\documentclass[12pt]{iopart}

\usepackage[bookmarks=true,colorlinks=true,hyperindex=true,linkcolor=red,citecolor=red]{hyperref} 
\usepackage[utf8]{inputenc}
\usepackage{bm}
\usepackage{graphicx}
\usepackage{color}
\usepackage[section]{placeins}
\usepackage{multicol}
\usepackage{enumerate}
\usepackage{ulem}
\usepackage{textcomp}
\usepackage{iopams}

\newcommand{\CCC}{{\mathcal C}}
\newcommand{\DDD}{{\mathcal D}}
\newcommand{\BBB}{{\mathcal B}}
\newcommand{\NNN}{{\mathcal N}}
\newcommand{\aaa}{{\mathbf a}}
\newcommand{\II}{{\mathbf I}}
\newcommand\bra[1]{\langle{#1}|}
\newcommand\ket[1]{|{#1}\rangle}

\newcommand{\bket}[1]{\left<#1\right>}
\newcommand{\abs}[1]{\left|#1\right|}

\newcommand{\ba}{\textrm{\bf{a}}}

\newcommand{\bb}{\textrm{\bf{b}}}
\newcommand{\bH}{\textrm{\bf{H}}}

\global\long\def\a{\alpha}
\global\long\def\b{\beta}
\global\long\def\g{\gamma}

\global\long\def\t{\theta}

\global\long\def\k{\kappa}

\global\long\def\m{\mu}

\global\long\def\r{\rho}

\global\long\def\S{\Sigma}

\global\long\def\P{\Phi}
\newcommand{\vbe}{\begin{equation}}
\renewcommand{\vee}{\end{equation}}
\newcommand{\vba}{\begin{eqnarray}}
\newcommand{\vea}{\end{eqnarray}}
\newcommand{\ke}{\rangle}
\newcommand{\cp}{|{\mathcal{C}_\alpha^{+}}\rangle}
\newcommand{\cm}{|{\mathcal{C}_\alpha^{-}}\rangle}
\newcommand{\cpb}{\langle{\mathcal{C}_\alpha^{+}}|}
\newcommand{\cmb}{\langle{\mathcal{C}_\alpha^{-}}|}
\newcommand{\rin}{\rho(0)}
\newcommand{\ph}{\aaa^\dag\aaa}
\newcommand{\half}{\frac{1}{2}}

\renewcommand{\tr}[1]{\textrm{tr}\{#1\}}
\newcommand{\rinf}{\rho_\infty}
\newcommand{\pht}{\textrm{\scriptsize ph}}
\newcommand{\modf}{\textrm{mod}4}
\newcommand{\td}[1]{\dot{#1}}

\newcommand{\moda}{|\a|}
\newcommand{\modas}{|\a|^2}
\newcommand{\modb}{|\b|}
\newcommand{\modbs}{|\b|^2}

\begin{document}

\title[]{Dynamically protected cat-qubits:\\ a new paradigm for universal quantum computation}

\author{Mazyar Mirrahimi$^{1,2}$, Zaki Leghtas$^2$, Victor V. Albert$^{2,3}$, Steven Touzard$^2$, Robert J. Schoelkopf$^{2,3}$, Liang Jiang$^{2,3}$, Michel H. Devoret$^{2,3}$}
\address{
$^1$ INRIA Paris-Rocquencourt, Domaine de Voluceau, B.P. 105, 78153 Le Chesnay Cedex, France\\
$^2$ Department of Applied Physics, Yale University, New Haven, Connecticut 06520, USA\\
$^3$ Department of Physics, Yale University, New Haven, Connecticut 06520, USA\\
}
\ead{mazyar.mirrahimi@inria.fr}
\begin{abstract}
We present a new hardware-efficient paradigm for universal quantum computation which is based on encoding, protecting and manipulating quantum information in a quantum harmonic oscillator. This proposal exploits multi-photon driven dissipative processes to encode quantum information in logical bases composed of Schr\"odinger cat states. More precisely, we consider two schemes. In a first scheme, a two-photon driven dissipative process is used to stabilize a logical qubit basis of two-component Schr\"odinger cat states. While such a scheme ensures a protection of the logical qubit against the photon dephasing errors, the prominent error channel of single-photon loss induces bit-flip type errors  that cannot be corrected. Therefore, we consider a second scheme based on a four-photon driven dissipative process which leads to the choice of four-component Schr\"odinger cat states as the logical qubit. Such a logical qubit can be protected against single-photon loss  by continuous photon number parity measurements. Next, applying some specific  Hamiltonians, we provide a set of universal quantum gates on the encoded qubits of each of the two schemes. In particular, we illustrate how these operations can be rendered fault-tolerant with respect to various decoherence channels of participating quantum systems. Finally, we also propose experimental schemes based on quantum superconducting circuits and inspired by methods used in Josephson parametric amplification, which should allow to achieve these driven dissipative processes along with the Hamiltonians ensuring the universal operations in an efficient manner. 
\end{abstract}

\maketitle

\section{Introduction}\label{sec:intro}
In a recent paper~\cite{Leghtas-al-PRL-2013}, we showed that a quantum harmonic oscillator could be used as a powerful resource to encode and protect  quantum information. In contrast to the usual approach of multi-qubit quantum error correcting codes~\cite{Shor-QEC,Steane-PRL_1996}, we benefit from the infinite dimensional Hilbert space of a quantum harmonic oscillator to encode redundantly  quantum information while no extra decay channels are added. Indeed, the far dominant decay channel for a quantum harmonic oscillator, for instance, a microwave cavity field mode, is  photon loss. Hence, we only need one type of error syndrome to identify the photon loss error.  In this paper, we aim to extend the proposal of~\cite{Leghtas-al-PRL-2013} as a hardware-efficient protected quantum memory towards a hardware-efficient protected logical qubit with which we can perform universal quantum computations~\cite{nielsen-chang-book}. 

Before getting to this extension, we recall the idea behind the proposal of~\cite{Leghtas-al-PRL-2013}. We start by mapping the qubit state $c_0\ket{0}+c_1\ket{1}$ into a multi-component superposition of coherent states of the harmonic oscillator $\ket{\psi_\alpha^{(0)}}=c_0\ket{0}_L+c_1\ket{1}_L=c_0\ket{\CCC_\alpha^+}+c_1\ket{\CCC_{i\alpha}^+}$, where
$$
\ket{\CCC_\alpha^\pm}=\NNN(\ket{\alpha}\pm\ket{-\alpha}),\qquad \ket{\CCC_{i\alpha}^\pm}=\NNN(\ket{i\alpha}\pm\ket{-i\alpha}).
$$
Here, $\NNN (\approx 1/\sqrt 2)$ is a normalization factor, and $\ket{\alpha}$ denotes a coherent state of complex amplitude $\alpha$. By taking $\alpha$ large enough, $\ket{\alpha}$, $\ket{-\alpha}$, $\ket{i\alpha}$ and $\ket{-i\alpha}$ are quasi-orthogonal (note that for $\alpha=2$ considered in most simulations of this paper, $|\langle \alpha|i\alpha\rangle\rangle|^2<10^{-3}$). 
Such an encoding protects the quantum information against  photon loss events. In order to see this, let us  also define $\ket{\psi_\alpha^{(1)}}=c_0\ket{\CCC_{\alpha}^-}+ic_1\ket{\CCC_{i\alpha}^-}$, $\ket{\psi_\alpha^{(2)}}=c_0\ket{\CCC_\alpha^+}-c_1\ket{\CCC_{i\alpha}^+}$ and $\ket{\psi_\alpha^{(3)}}=c_0\ket{\CCC_\alpha^-}-ic_1\ket{\CCC_{i\alpha}^-}$. The state $\ket{\psi_\alpha^{(n)}}$ evolves after a photon loss event to $\aaa\ket{\psi_\alpha^{(n)}}/\|\aaa\ket{\psi_\alpha^{(n)}}\|=\ket{\psi_\alpha^{[(n+1)\textrm{\scriptsize mod}4]}}$, where $\aaa$ is the harmonic oscillator's annihilation operator. Furthermore, in the absence of jumps during a time interval $t$, $\ket{\psi_\alpha^{(n)}}$ deterministically evolves to $\ket{\psi_{\alpha e^{-\kappa t/2}}^{(n)}}$, where $\kappa$ is the decay rate of the harmonic oscillator. Now, the parity operator $\Pi=\exp(i\pi \aaa^\dag\aaa)$ can act as a photon jump indicator. Indeed, we have  $\langle\psi_\alpha^{(n)}~|~\Pi~|~\psi_\alpha^{(n)}\rangle=(-1)^n$ and therefore the measurement of the photon number parity can indicate the occurrence of a photon loss event.  While the parity measurements keep track of the photon loss events, the deterministic relaxation of the energy, replacing $\alpha$ by $\alpha e^{-\kappa t/2}$ remains inevitable. To overcome this relaxation of energy, we need to intervene before the coherent states start to overlap in a significant manner to re-pump  energy into the codeword. 

In~\cite{Leghtas-al-PRL-2013}, applying some tools that were introduced in~\cite{Leghtas-al-PRA-2013}, we illustrated that simply coupling a cavity mode to a single superconducting qubit in the strong dispersive regime~\cite{schuster-nature07} provides the required controllability over the cavity mode (modeled as a quantum harmonic oscillator) to perform all the tasks of quantum information encoding, protection and energy re-pumping. The proposed tools exploit the fact that in such a coupling regime, both qubit and cavity frequencies split into well-resolved spectral lines indexed by the number of excitations in the qubit and the cavity. Such a splitting in the frequencies gives the possibility of performing operations controlling the joint qubit-cavity state. For instance, the energy re-pumping into the Schr\"odinger cat state is performed by decoding back the quantum information onto the physical qubit and re-encoding it on the cavity mode by re-adjusting the number of photons. However such an invasive control of the state  exposes the quantum information to decay channels (such as the $T_1$ and the $T_2$ decay processes of the physical qubit) and limits the performance of the protection scheme. Furthermore, if one wanted to use this quantum memory as a protected logical qubit, the application of quantum gates on the encoded information would require the decoding of this information onto the physical qubits, performing the operation, and re-encoding it back to the cavity mode. Once again, by exposing the quantum information to un-protected qubit decay channels, we limit the fidelity of these gates.

In this paper, we aim to exploit an engineered coupling of the storage cavity mode to its environment in order to maintain the energy of the encoded Schr\"odinger cat state. It is well-known that resonantly driving a damped quantum harmonic oscillator stabilizes a coherent state of the cavity mode field. In particular, the complex amplitude $\alpha$ of this coherent state depends linearly on the complex amplitude of the driving field. In contrast, it has been proposed that coupling a quantum harmonic oscillator to a bath where any energy exchange with the bath happens in pairs of photons, one can drive the quantum harmonic oscillator to the two aforementioned two-component Schr\"odinger cat states $\ket{\CCC_\alpha^+}$ and $\ket{\CCC_{\alpha}^-}$~\cite{Carmichael-Wolinsky-1988,Krippner-et-al-1994,Hach-Gerry-1994,Gilles-et-al-1994,Everitt-et-al-2013}. In Sec.~\ref{sec:pump} and~\ref{append:victor}, we will exploit such a two-photon driven dissipative process and extend the results of~\cite{Carmichael-Wolinsky-1988,Krippner-et-al-1994,Hach-Gerry-1994,Gilles-et-al-1994} by analytically determining the asymptotic behavior of the system for any initial state. In particular, we will illustrate how such a two-photon process can lead to take the Schr\"odinger cat states $\ket{\CCC_\alpha^+}$ and $\ket{\CCC_{\alpha}^-}$ (or equivalently the coherent states $\ket{\pm{\alpha}}$) as logical $\ket{0}$ and  $\ket{1}$ of a qubit which is protected against  a photon dephasing error channel. Such a logical qubit, however, is not protected against the dominant single-photon loss channel. Therefore, in the same section, we propose an extension of this two-photon process to a four-photon process for which the Schr\"{o}dinger cat states $\ket{\CCC_{\alpha}^{(0\textrm{\scriptsize mod}4)}}=\NNN(\ket{\CCC_{\alpha}^+}+\ket{\CCC_{i\alpha}^+})$ and $\ket{\CCC_{\alpha}^{(2\textrm{\scriptsize mod}4)}}=\NNN(\ket{\CCC_{\alpha}^+}-\ket{\CCC_{i\alpha}^+})$ (or equivalently the states $\ket{\CCC_\alpha^+}$ and $\ket{\CCC_{i\alpha}^+}$) become a natural choice of logical $\ket{0}$ and $\ket{1}$. Thus, we end up with a logical qubit which is protected against photon dephasing errors and for which we can also track and correct errors due to the dominant single-photon loss channel by continuous photon number parity measurements~\cite{Leghtas-al-PRL-2013}. 

In Sec.~\ref{sec:gates}, we present a toolbox to perform universal quantum computation with such protected Schr\"odinger cat states~\cite{Gilchrist-et-al-2004}. Applying specific  Hamiltonians that should be  easily engineered by methods inspired by those used in Josephson parametric amplification, and in the presence of the two-photon or four-photon driven dissipative processes, we can very efficiently perform operations such as arbitrary rotations around the Bloch sphere's $X$ axis and a two-qubit entangling gate.  These schemes can be well understood through quantum Zeno dynamics~\cite{Facchi-Pascazio-2002,raimond-et-al-2010,Raimond-et-al-2012} where the strong two-photon or four-photon processes project the evolution onto the degenerate subspace of the logical qubit (also known as a decoherence-free subspace~\cite{Lidar-et-al-98}). In order to achieve a full set of universal gates, we then only need to perform a $\pi/2$-rotation around the Bloch sphere's $Y$ or $Z$ axis. This is performed by the Kerr effect, induced  when we couple the cavity mode to a nonlinear medium such as a Josephson junction~\cite{Yurke-Stoler-1986,Kirchmair-al-Nature_2013}. We will illustrate that these gates remain protected against the decay channels of all involved quantum systems and could therefore be employed in a fault-tolerant quantum computation protocol. 

Finally, in Sec.~\ref{sec:realization}, we propose a readily realizable experimental scheme to achieve the two-photon driven dissipative process along with Hamiltonians needed for universal logical gates.  Indeed, we will illustrate that a simple experimental design based on circuit  quantum electrodynamics gives us enough flexibility to engineer all the Hamiltonians and the damping operator that are required for the protocols related to the two-photon process. Focusing on a fixed experimental setup, we will only need to  apply different pumping drives of well-chosen but fixed amplitudes and frequencies to achieve these requirements. Moreover, comparing to the experimental scheme proposed in~\cite{Everitt-et-al-2013} (based on the proposal by~\cite{Kumar-Divincenzo-2010}) our scheme does not require any symmetries in hardware design: in particular, the frequencies of the modes involved in the hardware could be very different, which helps to achieve an important separation of decay times for the two modes. As supporting indications, similar devices with parameters close to those required in this paper have been recently realized and characterized experimentally~\cite{Kirchmair-al-Nature_2013,Vlastakis-et-al-Science_2013}.  An extension of this experimental scheme to the case of the four-photon driven dissipative process is currently under investigation and we will describe the starting ideas.

\section{Driven dissipative multi-photon processes and protected logical qubits}\label{sec:pump}

\begin{figure}[h]
\begin{center}
\includegraphics[width=\textwidth]{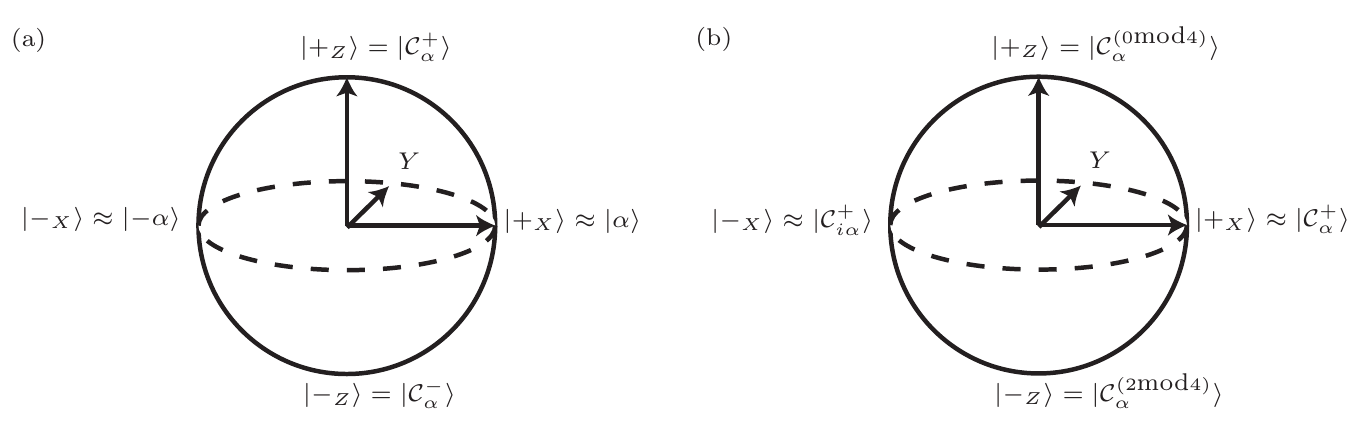}
\end{center}
\caption{\textbf{(a)} The two-photon driven dissipative process leads to the choice of even  and  odd Schr\"{o}dinger cat states $\ket{\CCC_\alpha^+}$ and $\ket{\CCC^\alpha_-}$ as the logical $\ket{0}$ and $\ket{1}$ of a qubit not protected against the single-photon loss channel. In this encoding, the $\ket{+_X}$ and $\ket{-_X}$ Bloch vectors approximately correspond to the coherent states $\ket{\alpha}$ and $\ket{-\alpha}$ (the approximate correspondence is due to the non-orthogonality of the two coherent states which is suppressed exponentially by $4|\alpha|^2$; While the coherent states are quasi-orthogonal, the cat states are orthogonal for all values of $\a$; Since the overlap between coherent states decreases exponentially with $\modas$, the two sets of states can be considered as approximately mutually unbiased bases for an effective qubit for $\moda \gtrsim2$). \textbf{(b)} The four-photon driven dissipative process leads to the choice of four-component Schr\"{o}dinger cat states $\ket{\CCC_\alpha^{(0\textrm{\scriptsize mod}4)}}$ and $\ket{\CCC_\alpha^{(2\textrm{\scriptsize mod}4)}}$ as the logical $\ket{0}$ and $\ket{1}$ of a qubit which can be protected against single-photon loss channel by continuous photon number parity measurements. Here $\ket{\CCC_\alpha^{(0\textrm{\scriptsize mod}4)}}=\NNN(\ket{\CCC_\alpha^+}+\ket{\CCC_{i\alpha}^+})$ corresponds to a 4-cat state which in the Fock basis is only composed of photon number states that are multiples of four. Similarly $\ket{\CCC_\alpha^{(2\textrm{\scriptsize mod}4)}}=\NNN(\ket{\CCC_\alpha^+}-\ket{\CCC_{i\alpha}^+})$ corresponds to a 4-cat state which in the Fock basis is composed of states whose photon numbers are the even integers not multiples of 4. In this encoding, $\ket{+_X}$ and $\ket{-_X}$ Bloch vectors approximately correspond to the two-component Schr\"{o}dinger cat states $\ket{\CCC_\alpha^{+}}$ and $\ket{\CCC_\alpha^{-}}$.}
\label{fig:Bloch}
\end{figure}

\subsection{Two-photon driven dissipative process}\label{ssec:twoPh}
\begin{figure}[h]
\begin{center}
\includegraphics[width=\textwidth]{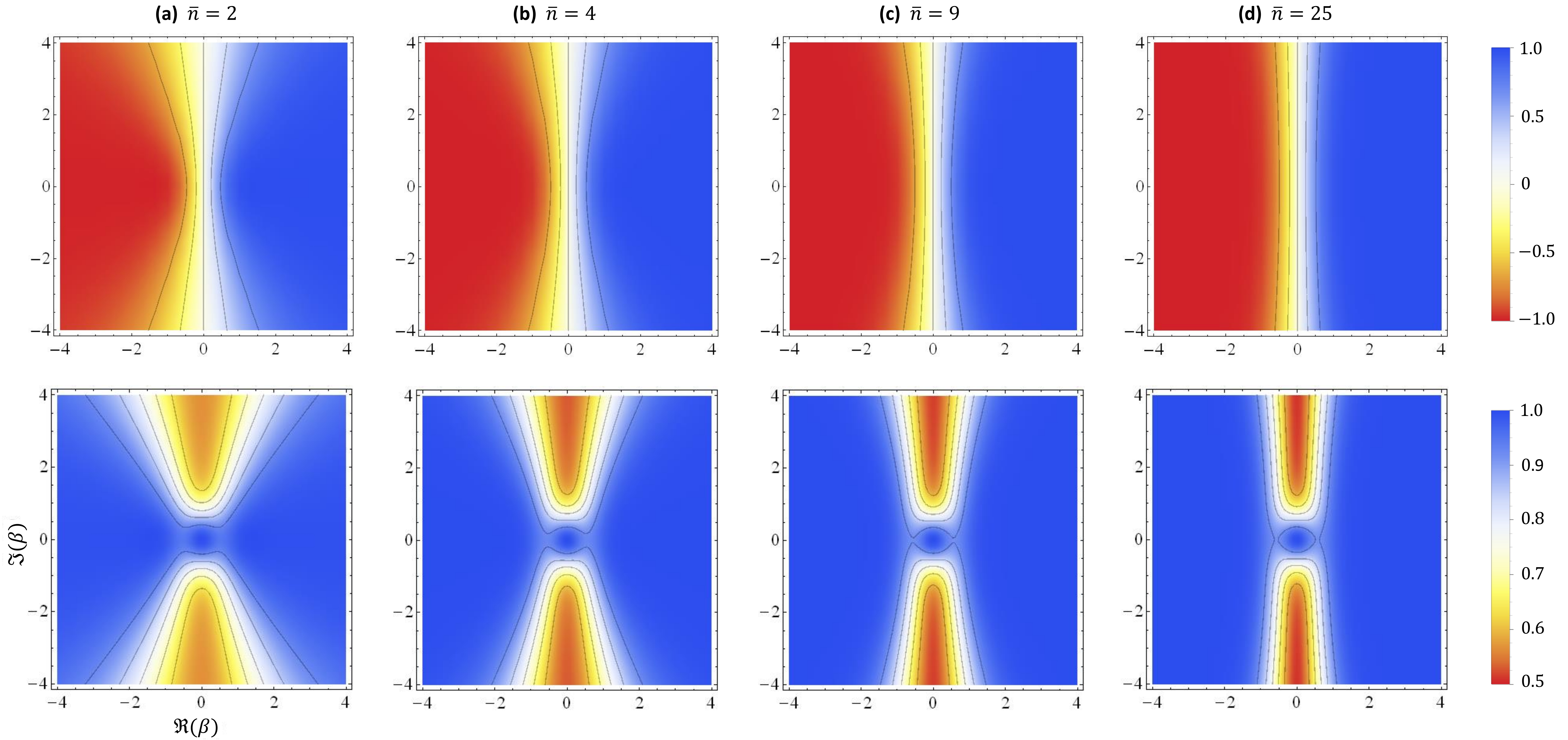}
\end{center}
\caption{Asymptotic (infinite-time) behavior of the two-photon driven dissipative process given by Eq.~(\ref{eq:Two-Photon}) where the density matrix is initialized in a coherent state. Here a point $\beta$ in the phase space corresponds to the coherent state $\ket{\beta}$ at which the process is initialized. The upper row illustrates the value of the Bloch sphere $X$-coordinate in the logical basis $\{\ket{\CCC_\alpha^+},\ket{\CCC_\alpha^-}\}$ ($\approx\bket{\alpha|\rho_s|\alpha}-\bket{-\alpha|\rho_s|-\alpha}$) where $\alpha=\sqrt{\bar n}=\sqrt{2\epsilon_{2\textrm{\scriptsize ph}}/\kappa_{2\textrm{\scriptsize ph}}}$ for $\bar n=2, 4, 9$ and $25$. We observe that for most coherent states except for a narrow vertical region in the center of the phase space, the system converges to one of the steady coherent states $\ket{\pm\alpha}$. The lower row illustrates the purity of the steady state to which we converge ($\tr{\rho_\infty^2}$) for various initial coherent states. Besides the asymptotic state being the pure $\ket{\pm\alpha}$ away from the vertical axis, one can observe that the asymptotic state is also pure for initial states near the center of phase space. Indeed, starting in the vacuum state, the two-photon process drives the system to the pure  Schr\"odinger cat state $\ket{\CCC_\alpha^+}$.
}  
\label{fig:charac_2ph}
\end{figure}
Let us consider the harmonic oscillator to be initialized in the vacuum state and let us drive it by an external field in such a way that it can only absorb photons in pairs. Assuming furthermore that the energy decay also only happens in pairs of photons, one easily observes that the photon number parity is conserved. More precisely, we consider the master equation corresponding to a two-photon driven dissipative quantum harmonic oscillator (with $\td{\r}$ being the time derivative of $\r$)
\begin{eqnarray}\label{eq:Two-Photon}
\td{\rho}=[\epsilon_{2\textrm{\scriptsize ph}}\aaa^{\dag 2}-\epsilon_{2\textrm{\scriptsize ph}}^*\aaa^2,\rho]+\kappa_{2\textrm{\scriptsize ph}}\DDD[\aaa^2]\rho,
\end{eqnarray}
where 
$$
\DDD[A]\rho=A\rho A^\dag-\frac{1}{2} A^\dag A\rho-\frac{1}{2}\rho A^\dag A.
$$
When $\rho(0)=\ket{0}\bra{0}$, one can show that the density matrix $\rho$ converges towards a pure even Schr\"odinger cat state given by the wavefunction $\ket{\CCC_\alpha^+}=\NNN(\ket{\alpha}+\ket{-\alpha})$, where $\alpha=\sqrt{2\epsilon_{2\textrm{\scriptsize ph}}/\kappa_{2\textrm{\scriptsize ph}}} $ and $\NNN$ is a normalizing factor. Similarly, if the system is initiated in a state with an odd photon number parity such as the Fock state $\ket{1}\bra{1}$, it converges towards the pure odd Schr\"odinger cat state $\ket{\CCC_\alpha^-}=\NNN(\ket{\alpha}-\ket{-\alpha})$. Indeed, the set of steady states of Eq.~(\ref{eq:Two-Photon}) is given by the  set of density operators defined on the two dimensional Hilbert space spanned by $\{\ket{\alpha},\ket{-\alpha}\}$~\cite{Gilles-et-al-1994}.  For any initial state, the system exponentially converges to this set in infinite time, making the span of $\{\ket{-\alpha},\ket{\alpha}\}$ the asymptotically stable manifold of the system. However, the asymptotic states in this manifold are not always pure states.  One of the results of this paper is to characterize the asymptotic behavior of the above dynamics for any  initial state (see~\ref{append:victor}). In particular, initializing the system in a coherent state $\rho(0)=\ket{\beta}\bra{\beta}$, it converges to the steady state 
\begin{equation}\label{eq:rinf}
\rho_\infty=c_{++}\ket{\CCC_\alpha^+}\bra{\CCC_\alpha^+}+c_{--}\ket{\CCC_\alpha^-}\bra{\CCC_\alpha^-}+c_{+-}\ket{\CCC_\alpha^+}\bra{\CCC_\alpha^-}+c_{+-}^*\ket{\CCC_\alpha^-}\bra{\CCC_\alpha^+},
\end{equation}
with
\begin{eqnarray*}
c_{++}&=\frac{1}{2}\left(1+e^{-2|\beta|^2}\right),\qquad c_{--}=\frac{1}{2}\left(1-e^{-2|\beta|^2}\right),\\
c_{+-}&=\frac{i\alpha \beta^* e^{-|\beta|^2}}{\sqrt{2\sinh(2|\alpha|^2)}}\int_{\phi=0}^\pi d\phi e^{-i\phi} I_0(|\alpha^2-\beta^2e^{2i\phi}|),
\end{eqnarray*}
where $I_0(.)$ is the modified Bessel function of the first kind. For large enough $|\beta|$, the populations of the even and odd cat states $\ket{\CCC_\alpha^\pm}$, $c_{++}$ and $c_{--}$  respectively, equilibrate to one-half. At large enough $\alpha$ (see Fig.~\ref{fig:charac_2ph} top row), if one initializes with a coherent state away from the vertical axis in phase space, then the system will converge towards one of the two steady coherent states $\ket{\pm\alpha}$ (with the sign depending on whether one initialized to the right or the left of the vertical axis). This suggests that if we choose the states $\ket{\CCC_\alpha^+}$ and $\ket{\CCC_\alpha^-}$ as the logical qubit states (see Fig.~\ref{fig:Bloch}(a)), the two Bloch vectors $\ket{+_X}\approx\ket{\alpha}$ and $\ket{-_X}\approx\ket{-\alpha}$ are robustly conserved.  Therefore, we will deal with a qubit where the phase-flip errors are very efficiently suppressed and the dominant error channel is the bit-flip errors  (which could be induced by a single photon decay process). This could be better understood if we consider the presence of a dephasing error channel for the quantum harmonic oscillator. In the presence of dephasing with rate $\kappa_\phi$, but no single-photon decay (we will discuss this later), the  master equation of the driven system is given as follows 
\begin{eqnarray}\label{eq:deph}
\td{\rho}=[\epsilon_{2\textrm{\scriptsize ph}}\aaa^{\dag 2}-\epsilon_{2\textrm{\scriptsize ph}}^*\aaa^2,\rho]+\kappa_{2\textrm{\scriptsize ph}}\DDD[\aaa^2]\rho+\kappa_\phi\DDD[\aaa^\dag\aaa]\rho.
\end{eqnarray} 
Such a dephasing, similar to the photon drive and dissipation, does not affect the photon number parity. Therefore the populations of the cat states $\ket{\CCC_\alpha^+}$ and $\ket{\CCC_{\alpha}^-}$, or equivalently the $\ket{+_Z}$ and $\ket{-_Z}$ states in the logical basis, remain constant in the presence of such dephasing. This means that such an error channel does not induce any bit flip errors on the logical qubit. It can however induce phase flip errors. But as shown in the~\ref{append:victor}, the rate at which such logical phase flip errors happen is exponentially suppressed by the size of the cat.  Indeed, for $\kappa_\phi\ll \kappa_{2\textrm{\scriptsize ph}}$, the induced logical phase flip rate is given by
$$
\gamma_{\textrm{\scriptsize phase-flip}}\approx \kappa_\phi\frac{|\alpha|^2}{2\sinh(2|\alpha|^2)}\rightarrow 0 \quad\textrm{ as }\quad |\alpha|\rightarrow\infty.
$$
The two-photon driven dissipative process therefore leads to a logical qubit basis which is very efficiently protected against the harmonic oscillator's dephasing channel. It is, however, well known that the major decay channel in usual practical quantum harmonic oscillators is single-photon loss~\cite{haroche-raimond:book06}. While the two-photon process fixes the manifold spanned by the states $\ket{\CCC_\alpha^\pm}$ as the steady state manifold, the single-photon jumps, that can be modeled by application at a random time of the annihilation operator $\aaa$, lead to a bit-flip error channel on this logical qubit basis. Indeed, the application of $\aaa$ on $\ket{\CCC_\alpha^\pm}$ sends the state to $\ket{\CCC_\alpha^\mp}$. Such jumps are not suppressed by the two-photon  process and a single-photon decay rate of $\kappa_{1\textrm{\scriptsize ph}}$ leads to a logical qubit bit-flip rate of $|\alpha|^2\kappa_{1\textrm{\scriptsize ph}}$. It is precisely for this reason that we need to get back to the protocol of~\cite{Leghtas-al-PRL-2013} recalled in Sec.~\ref{sec:intro}. 

\subsection{Four-photon driven dissipative process}
In order to be able to track single-photon jump events, we need to replace the logical qubit states $\ket{\CCC_\alpha^\pm}$ by the Schr\"odinger cat states $\ket{\CCC_\alpha^{(0\textrm{\scriptsize mod}4)}}$ and $\ket{\CCC_\alpha^{(2\textrm{\scriptsize mod}4)}}$. To this aim, we present here an extension of the above two-photon process  to a four-photon one. Indeed, coupling a quantum harmonic oscillator to a driven bath in such a way that any exchange of energy with the bath happens through quadruples of photons, we get the following master equation:
\begin{eqnarray}\label{eq:Four-Photon}
\td{\rho}=[\epsilon_{4\textrm{\scriptsize ph}}\aaa^{\dag 4}-\epsilon_{4\textrm{\scriptsize ph}}^*\aaa^4,\rho]+\kappa_{4\textrm{\scriptsize ph}}\DDD[\aaa^4]\rho.
\end{eqnarray}
The steady states of these dynamics are given by the set of density operators defined on the 4-dimensional Hilbert space spanned by $\{\ket{\pm\alpha},\ket{\pm i\alpha}\}$ where $\alpha=(2\epsilon_{4\textrm{\scriptsize ph}}/\kappa_{4\textrm{\scriptsize ph}})^{1/4}$. In particular, noting that the above master equation conserves the number of photons modulo 4, starting at initial Fock states $\ket{0}$, $\ket{1}$, $\ket{2}$ and $\ket{3}$, the system converges, respectively, to the pure states 
\begin{eqnarray*}
\CCC_\alpha^{(0\textrm{\scriptsize mod}4)}&=\NNN(\ket{\CCC_\alpha^+}+\ket{\CCC_{i\alpha}^+}),\quad \CCC_\alpha^{(1\textrm{\scriptsize mod}4)}=\NNN(\ket{\CCC_\alpha^-}-i\ket{\CCC_{i\alpha}^-}),\\
\CCC_\alpha^{(2\textrm{\scriptsize mod}4)}&=\NNN(\ket{\CCC_\alpha^+}-\ket{\CCC_{i\alpha}^+}),\quad \CCC_\alpha^{(3\textrm{\scriptsize mod}4)}=\NNN(\ket{\CCC_\alpha^-}+i\ket{\CCC_{i\alpha}^-}).
\end{eqnarray*}
By keeping track of the photon number parity, we can restrict the dynamics to the even parity states, so that the steady states are given by the set of density operators defined on the Hilbert space spanned by $\{\ket{\CCC_{\alpha}^{(0\textrm{\scriptsize mod}4)}},\ket{\CCC_{\alpha}^{(2\textrm{\scriptsize mod}4)}}\}$. Similar to the two-photon process, these two states will be considered as the logical, now also protected, $\ket{0}$ and $\ket{1}$ of a qubit (see Fig.~\ref{fig:Bloch}(b)).  Once again, a photon dephasing channel of rate $\kappa_\phi$ leads to a phase-flip error channel for the logical qubit where the error rate is exponentially suppressed by the size of the Schr\"odinger cat state (see numerical simulations of~\ref{append:victor}). 

{Note that probing the photon number parity of a quantum harmonic oscillator in a quantum non-demolition manner can be performed by a Ramsey-type experiment where the cavity mode is dispersively coupled to a single qubit playing the role of the meter~\cite{haroche-et-al-2007}. Such an efficient continuous monitoring of the photon number parity has recently been achieved using a transmon qubit coupled to a 3D cavity mode in the strong dispersive regime~\cite{Sun-et-al-2013}. Furthermore, we have determined that this photon number parity measurement can be performed in a fault-tolerant manner; the encoded state can remain intact in the presence of various decay channels of the meter. The details of such a fault-tolerant parity measurement method will be addressed in a future publication~\cite{Herviou-Mirrahimi}.}

In summary, we have shown that one can achieve a logical qubit basis of cat states $\{\ket{\CCC_\alpha^+},\ket{\CCC_\alpha^-}\}$ through a two-photon driven dissipative process. A photon dephasing error channel is translated to a phase-flip error rate which is exponentially suppressed by the size of the cat states. A single-photon decay channel, however, leads to a bit-flip error channel whose rate is $|\alpha|^2$ times larger than the single-photon decay rate. In order to protect the qubit against such a prominent decay channel, we introduce the similar four-photon driven dissipative process whose logical qubit basis is given by the Schr\"odinger cat states $\{\ket{\CCC_\alpha^{(0\textrm{\scriptsize mod}4)}},\ket{\CCC_{\alpha}^{(2\textrm{\scriptsize mod}4)}}\}$. Once again, the photon dephasing error channel is replaced by a phase-flip error channel whose rate is suppressed exponentially by the size of the Schr\"odinger cat state. A single-photon decay channel leads to the transfer of quantum information to a new logical basis given by odd Schr\"odinger cat states $\{\ket{\CCC_{\alpha}^{(3\textrm{\scriptsize mod}4)}},\ket{\CCC_{\alpha}^{(1\textrm{\scriptsize mod}4)}}\}$. However, we can keep track of single photon decay by continuously measuring the photon number parity. Therefore, the cat-state logical qubit can be protected against single photon decay while also having photon dephasing errors exponentially suppressed. 
  
\section{Universal gates and fault-tolerance}\label{sec:gates}

The proposal of the previous section together with the implementation scheme of the next one should lead to a technically realizable protected quantum memory. Having discussed how one can dynamically protect from both bit-flip and phase-flip errors, we show in this section that such a protection scheme can  be further explored towards a new paradigm for performing fault-tolerant quantum computation. Having this in mind, we will show how a set of universal quantum gates can be efficiently implemented on such dynamically protected qubits. This set consists of arbitrary rotations around the $X$ axis of a single qubit, a single-qubit $\pi/2$ rotation around the $Z$ axis, and a two-qubit entangling gate. 

The arbitrary rotations around $X$-axis of a single qubit  and the two-qubit entangling gate can be generated by applying some fixed-amplitude driving fields at well-chosen frequencies, leading to additional terms in the effective Hamiltonian of the pumped regime. In order to complete this set of gates, one then only needs a single-qubit $\pi/2$-rotation around either the $Y$ or $Z$ axes. Here we perform such rotation around the $Z$ axis by turning off the multi-photon drives and  applying a Kerr effect in the Hamiltonian. Such a Kerr effect is naturally induced in the resonator mode through its coupling to the Josephson junction, providing the non-linearity needed for the multi-photon process.  Finally, we will also discuss the fault-tolerance properties of these gates. 

\subsection{Quantum Zeno dynamics for arbitrary rotations of a single qubit}
\label{sec:zenophaseshiftgate}
Let us start with the case of the two-photon process where the quantum information is not protected against single-photon loss. The parity eigenstates $\ket{\CCC_\alpha^+}$ and $\ket{\CCC_\alpha^-}$ are invariant states when the exchange of photons with the environment only happens through pairs of photons. Here, we are interested in performing a rotation of an arbitrary angle $\theta$ around the $X$ axis in this logical basis of $\{\ket{\CCC_{\alpha}^+},\ket{\CCC_{\alpha}^-}\}$:
$$
X_\theta=\cos\theta(\ket{\CCC_\alpha^+}\bra{\CCC_\alpha^+}+\ket{\CCC_\alpha^-}\bra{\CCC_\alpha^-})+i\sin\theta(\ket{\CCC_\alpha^+}\bra{\CCC_\alpha^-}+\ket{\CCC_\alpha^-}\bra{\CCC_\alpha^+}).
$$
In other to ensure such a population transfer between the even and odd parity manifolds, one can apply a Hamiltonian ensuring single-photon exchanges with the system. We show that the simplest Hamiltonian that ensures such a transfer of population is a driving field at resonance with the quantum harmonic oscillator. The idea consists of driving the quantum harmonic oscillator at resonance where the phase of the drive is chosen to be out of quadrature with respect to the Wigner fringes of the Schr\"odinger cat state.  Furthermore, the amplitude of the drive is chosen to be much smaller than the two-photon dissipation rate. This can be much better understood when reasoning in a time-discretized manner. Let us assume $\alpha$ to be real and  the quantum harmonic oscillator to be initialized in the even parity cat state $\ket{\CCC_\alpha^+}$. Applying a displacement operator $D(i\epsilon)=\exp(i\epsilon(\aaa+\aaa^\dag))$ with $\epsilon\ll 1$ brings the state towards
$$
D(i\epsilon)\ket{\CCC_\alpha^+}=\NNN(e^{-i\epsilon\alpha}\ket{-\alpha+i\epsilon}+e^{i\epsilon\alpha}\ket{\alpha+i\epsilon})
$$
Following the analysis of the previous section, the two-photon process re-projects this displaced state to the space spanned by $\{\ket{\CCC_\alpha^+},\ket{\CCC_\alpha^-}\}$ without significantly reducing the coherence term: the states $\ket{-\alpha+i\epsilon}$ and $\ket{\alpha+i\epsilon}$ are close to the coherent states $\ket{-\alpha}$ and $\ket{\alpha}$. Therefore, the displaced state is approximately projected on the state $\cos(\epsilon\alpha)\ket{\CCC_\alpha^+}+i\sin(\epsilon\alpha)\ket{\CCC_\alpha^-}$ . This is equivalent to applying an arbitrary rotation gate of the form $X_{\epsilon\alpha}$ on the initial cat state $\ket{\CCC_\alpha^+}$. This protocol can also be understood through quantum Zeno dynamics. The two-photon process can be thought of as a measurement which projects onto the steady-state space spanned by $\{\ket{\CCC_\alpha^+},\ket{\CCC_\alpha^-}\}$. Continuous performance of such a measurement freezes the dynamics in this space while the weak single-photon driving field ensures arbitrary rotations around $X$-axis of the logical qubit defined in this basis. 

\begin{figure}[h]
\begin{center}
\includegraphics[width=\textwidth]{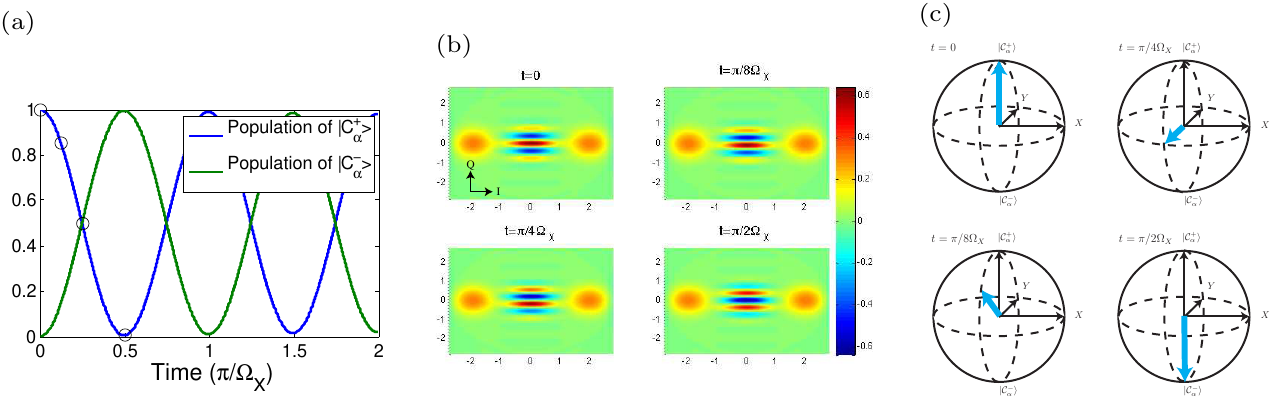}
\end{center}
\caption{Quantum Zeno dynamics as a tool for performing rotations of an arbitrary angle around the $X$-axis of the logical qubit space spanned by $\{\ket{\CCC_\alpha^+},\ket{\CCC_\alpha^-}\}$. The quantum harmonic oscillator is driven at resonance in the $Q$-direction while the two-photon driven dissipative process is acting on the system. \textbf{(a)} Simulations of Eq.~(\ref{eq:Zeno2Cat}) illustrate the Rabi oscillations around the Bloch sphere's $X$-axis in the logical qubit space at an effective Rabi frequency of $\Omega_X=2\epsilon_X\sqrt{\bar n}$. Here $\epsilon_X=\kappa_{2\textrm{\scriptsize ph}}/20$ and $\bar n=4$. \textbf{(b)} Wigner representation of the state at times $t=0$, $t=\pi/8\Omega_X$, $t=\pi/4\Omega_X$ and $t=\pi/2\Omega_X$. We can observe the shifts in the Wigner fringes while the state remains a coherent superposition with equal weights of $\ket{-\alpha}$ and $\ket{\alpha}$. \textbf{(c)} The tomography at these times  $t=0$, $t=\pi/8\Omega_X$, $t=\pi/4\Omega_X$ and $t=\pi/2\Omega_X$ illustrate  rotations of angles $0$, $\pi/4$, $\pi/2$ and $\pi$ around the logical $X$ axis.}
\label{fig:ArbPhase2Cat}
\end{figure}

In order to simulate such quantum Zeno dynamics, we consider the effective master equation
\begin{equation}\label{eq:Zeno2Cat}
 \td{\rho}=-i\epsilon_X[\aaa+\aaa^\dag,\rho]+\epsilon_{2\textrm{\scriptsize ph}}[\aaa^{\dag 2}-\aaa^2,\rho]+\kappa_{2\textrm{\scriptsize ph}}\DDD[\aaa^2]\rho.
\end{equation}
Here, taking $\epsilon_{2\textrm{\scriptsize ph}}=\bar n\kappa_{2\textrm{\scriptsize ph}}/2$ and $\epsilon_X\ll \kappa_{2\textrm{\scriptsize ph}}$, we ensure the above Zeno dynamics in the space spanned by $\{\ket{\CCC_\alpha^-},\ket{\CCC_\alpha^+}\}$, with $\alpha=\sqrt{\bar n}$ (here, the choice of the phase of $\epsilon_{2\textrm{\scriptsize ph}}$ fixes $\alpha$ to be real).  By initializing the system in the state $\ket{\CCC_\alpha^+}$ and letting the system evolve  following the above dynamics, we numerically simulate the equivalent of a Rabi oscillation's experiment. We monitor the population of the states $\ket{\CCC_\alpha^+}$ and $\ket{\CCC_\alpha^-}$ (the $\ket{+_Z}$ and $\ket{-_Z}$ states) during the evolution.  Fig.~\ref{fig:ArbPhase2Cat}(a) illustrates the result of such simulation over a time of $2\pi/\Omega_X$ where $\Omega_X$ the effective Rabi frequency is given by
$$
\Omega_X=2\epsilon_X\sqrt{\bar n}.
$$
This effective Rabi frequency can be found by projecting the added driving Hamiltonian $\epsilon_X(\aaa+\aaa^\dag)$ on the space spanned by $\{\ket{\CCC_\alpha^-},\ket{\CCC_\alpha^+}\}$:\small
$$
\epsilon_X\left(\Pi_{\ket{\CCC_\alpha^+}}+\Pi_{\ket{\CCC_\alpha^-}}\right)(\aaa+\aaa^\dag)\left(\Pi_{\ket{\CCC_\alpha^+}}+\Pi_{\ket{\CCC_\alpha^-}}\right)=(\alpha+\alpha^*)\epsilon_X\left(\ket{\CCC_\alpha^+}\bra{\CCC_\alpha^-}+\ket{\CCC_\alpha^-}\bra{\CCC_\alpha^+}\right)=\Omega_X \sigma_x^L,
$$\normalsize
where $\Pi_{\ket{\CCC_\alpha^\pm}}=\ket{\CCC_\alpha^\pm}\bra{\CCC_\alpha^\pm}$.
One can note in Fig.~\ref{fig:ArbPhase2Cat}(a) (where we have chosen $\epsilon_X=\kappa_{2\textrm{\scriptsize ph}}/20$), the slight decay of the Rabi oscillations as a function of time. This is due to the finite ratio $\kappa_{2\textrm{\scriptsize ph}}/\epsilon_X$, which adds higher order terms to the above effective dynamics. Indeed, similar computations to the one in~\ref{append:victor} can be performed to calculate the effective dephasing time due to these higher order terms. In practice, this induced decay can be reduced by choosing larger separation of time-scales (smaller $\epsilon_X/\kappa_{2\textrm{\scriptsize ph}}$)  at the expense of longer gate times. However, even a moderate factor of 20 ensures gate fidelities in excess of $99.5\%$.

As illustrated in Fig.~\ref{fig:ArbPhase2Cat}(b), we can calculate the Wigner function at particular times during the evolution. This is performed for the times $t=0$, $t=\pi/8\Omega_X$, $t=\pi/4\Omega_X$ and $t=\pi/2\Omega_X$ and, as illustrated in Fig.~\ref{fig:ArbPhase2Cat}(c), we observe rotations of angle 0, $\pi/4$, $\pi/2$ and $\pi$ around the logical $X$ axis for the qubit states defined as $\ket{\CCC_\alpha^+}$ and $\ket{\CCC_\alpha^-}$. 

\begin{figure}[h]
\begin{center}
\includegraphics[width=\textwidth]{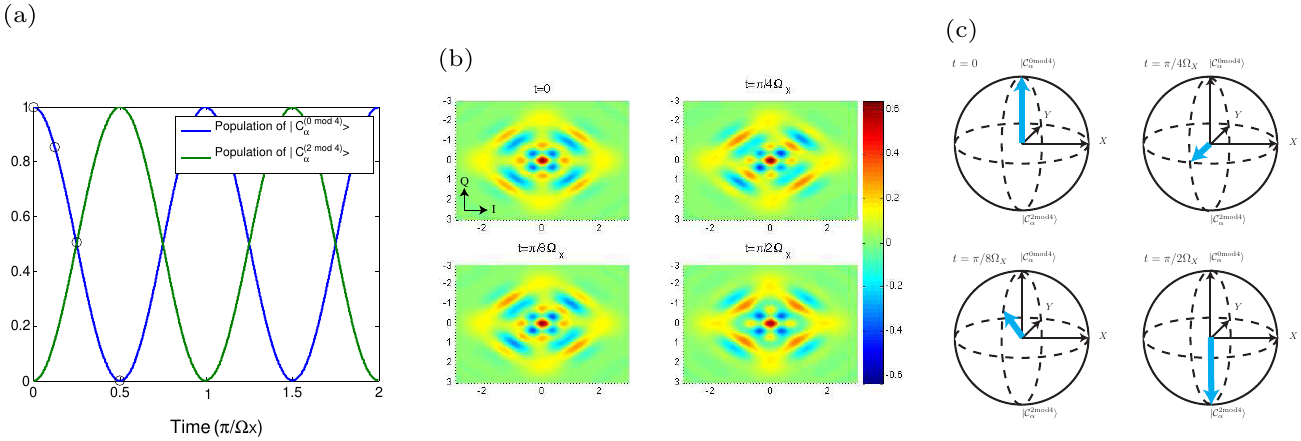}
\end{center}
\caption{Quantum Zeno dynamics as a tool for performing rotations of arbitrary angles around the Bloch sphere's $X$-axis of the logical qubit basis of $\{\ket{\CCC_\alpha^{(0\textrm{\scriptsize mod}4)}},\ket{\CCC_\alpha^{(2\textrm{\scriptsize mod}4)}}\}$.  A squeezing Hamiltonian is applied on the quantum harmonic oscillator while the four-photon driven dissipative process is acting. \textbf{(a)} Rabi oscillations around the $X$ axis with an effective Rabi frequency of $\Omega_X=2\epsilon_X \bar n$. Here $\epsilon_X=\kappa_{2\textrm{\scriptsize ph}}/20$ and $\bar n=4$. \textbf{(b)} Wigner representation of the state at times $t=0$, $t=\pi/8\Omega_Z$, $t=\pi/4\Omega_Z$ and $t=\pi/2\Omega_Z$, with different fringe patterns associated to rotations with different angles. \textbf{(c)} The tomography at these times  $t=0$, $t=\pi/8\Omega_Z$, $t=\pi/4\Omega_Z$ and $t=\pi/2\Omega_Z$ illustrate  rotations of angles $0$, $\pi/4$, $\pi/2$ and $\pi$ around the logical $X$ axis.}
\label{fig:ArbPhase4Cat}
\end{figure}

Let us now extend this idea to the case of the four-photon process where  quantum information can be protected through continuous parity measurements. For the two-photon process, a population transfer from the even cat state $\ket{\CCC_\alpha^+}$ to $\ket{\CCC_\alpha^-}$, is ensured through a resonant drive ensuring single-photon exchanges with the system. For the four-photon case, such a rotation of an arbitrary angle around the Bloch sphere's $X$-axis necessitates a population transfer between the two states  $\ket{\CCC_{\alpha}^{0\textrm{\scriptsize mod}4}}$ and $\ket{\CCC_{\alpha}^{2\textrm{\scriptsize mod}4}}$. The state $\ket{\CCC_{\alpha}^{0\textrm{\scriptsize mod}4}}$ correspond to a four-component Schr\"odinger cat state which in the Fock basis is only composed of states with photon numbers that are multiples of 4. Similarly, the state $\ket{\CCC_{\alpha}^{2\textrm{\scriptsize mod}4}}$ corresponds to a four-component Schr\"odinger cat state which in the Fock basis is only composed of photon number states that are even but not multiples of 4. Therefore, in order to ensure a population transfer from $\ket{\CCC_{\alpha}^{(0\textrm{\scriptsize mod}4)}}$ to $\ket{\CCC_{\alpha}^{(2\textrm{\scriptsize mod}4)}}$, we need to apply a Hamiltonian that adds/subtracts pairs of photons to/from the system. This can be done by adding a squeezing Hamiltonian of the form $\epsilon_X(e^{i\phi}a^2+e^{-i\phi}a^{\dag 2})$ to the Hamiltonian of the four-photon process (for a real $\alpha$, we take $\phi=0$ in order to be in correct quadrature with respect to the  Wigner fringes): 
\begin{equation}\label{eq:Zeno4Cat}
\td{\rho}=-i\epsilon_X[\aaa^2+\aaa^{\dag 2},\rho]+\epsilon_{4\textrm{\scriptsize ph}}[\aaa^{\dag 4}-\aaa^4,\rho]+\kappa_{4\textrm{\scriptsize ph}}\DDD[\aaa^4]\rho.
\end{equation}
In direct correspondence with the two-photon process, we initialize the system in the state $\ket{\CCC_\alpha^{(0\textrm{\scriptsize mod}4)}}$ and we simulate Eq.~(\ref{eq:Zeno4Cat}). Here $\epsilon_{4\textrm{\scriptsize ph}}=\bar n^2\kappa_{4\textrm{\scriptsize ph}}/2$ ensures that the subspace spanned by $\{\ket{\alpha},\ket{-\alpha},\ket{i\alpha},\ket{-i\alpha}\}$, with $\alpha=\sqrt{\bar  n}$ is asymptotically stable. Since all the Hamiltonians and decay terms correspond to exchanges of photons in pairs or quadruples and since we have initialized in  $\ket{\CCC_\alpha^{0\textrm{\scriptsize mod}4}}$, we can restrict the dynamics to the subspace spanned by even Fock states. In this subspace, the asymptotic manifold is generated by $\ket{\CCC_\alpha^{(0\textrm{\scriptsize mod}4)}}$ and $\ket{\CCC_\alpha^{(2\textrm{\scriptsize mod}4)}}$. We also take $\epsilon_X$ to be much smaller than $\kappa_{4\textrm{\scriptsize ph}}$. Simulations shown in Fig.~\ref{fig:ArbPhase4Cat}(a) (for $\bar n=4$ and $\epsilon_X=\kappa_{4\textrm{\scriptsize ph}}/20$) illustrate the Rabi oscillations at frequency 
$$
\Omega_X=2\epsilon_X \bar n
$$ 
around the Bloch sphere's $X$ axis in this logical basis. This Rabi frequency can also be retrieved by projecting the squeezing Hamiltonian onto the qubit subspace: 
\begin{eqnarray*}
&\fl  \epsilon_X \left(\Pi_{\ket{\CCC_\alpha^{(0\textrm{\scriptsize mod}4)}}}+\Pi_{\ket{\CCC_\alpha^{(2\textrm{\scriptsize mod}4)}}}\right)(\aaa^2+\aaa^{\dag 2})\left(\Pi_{\ket{\CCC_\alpha^{(0\textrm{\scriptsize mod}4)}}}+\Pi_{\ket{\CCC_\alpha^{(2\textrm{\scriptsize mod}4)}}}\right)=\\
&\qquad(\alpha^2+\alpha^{*2})\epsilon_X\left(\ket{\CCC_\alpha^{(0\textrm{\scriptsize mod}4)}}\bra{\CCC_\alpha^{(2\textrm{\scriptsize mod}4)}}+\ket{\CCC_\alpha^{(2\textrm{\scriptsize mod}4)}}\bra{\CCC_\alpha^{(0\textrm{\scriptsize mod}4)}}\right)=\Omega_X \sigma_x^L.
\end{eqnarray*}
As shown in Figures~\ref{fig:ArbPhase4Cat}(b),(c), we efficiently achieve an effective single-qubit gate corresponding to rotations of an arbitrary angle around the Bloch sphere's $X$-axis for the logical qubit spanned by $\{\ket{\CCC_\alpha^{(0\textrm{\scriptsize mod}4)}},\ket{\CCC_\alpha^{(2\textrm{\scriptsize mod}4)}}\}$.

\subsection{Quantum Zeno dynamics for a two-qubit entangling gate}
\label{sec:zenotwoqubitentanglement}
\begin{figure}[h]
\begin{center}
\includegraphics[width=\textwidth]{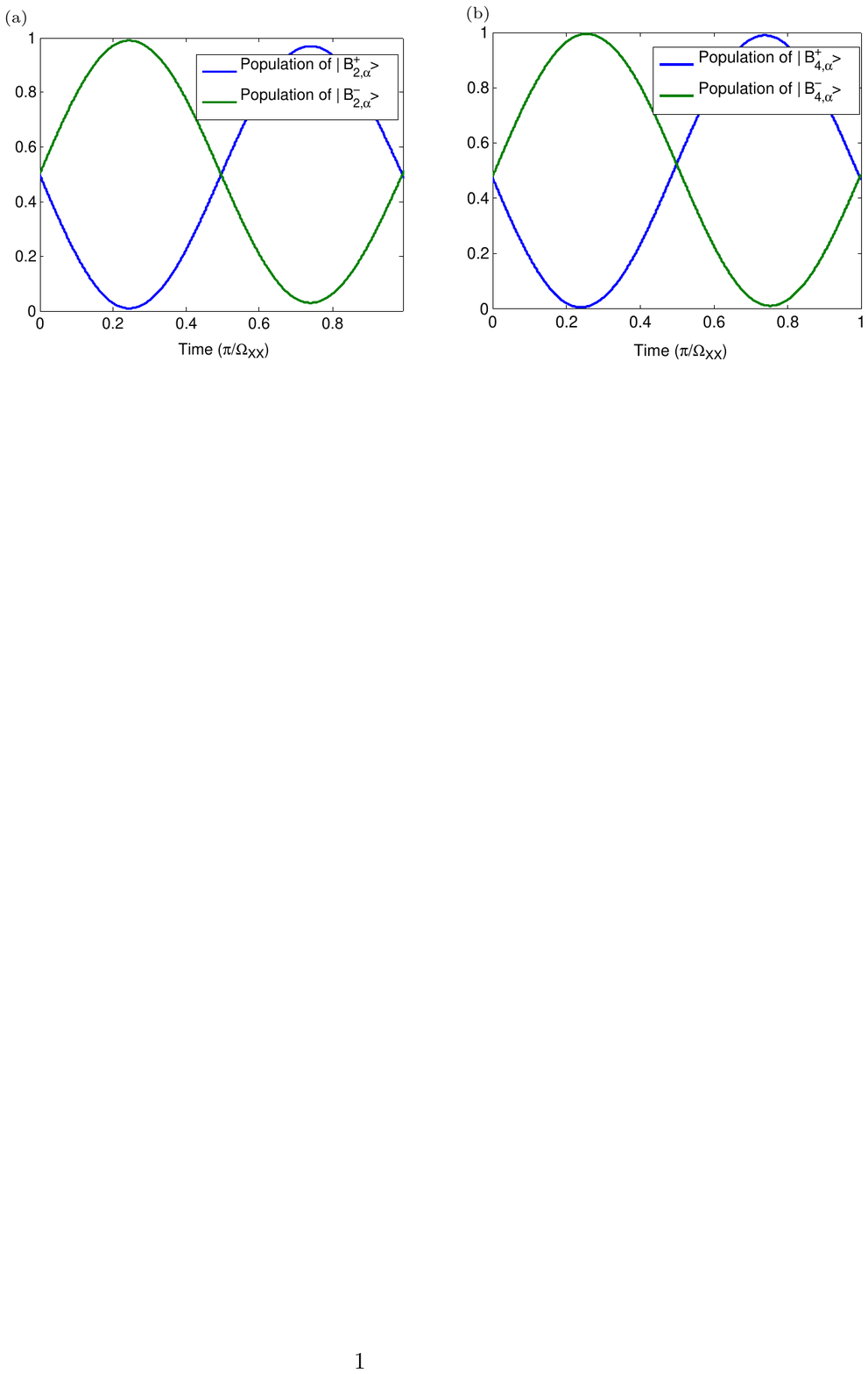}
\end{center}
\caption{{Quantum Zeno dynamics as a tool for performing a two-qubit entangling gate for the two cases of the two-photon process, with the logical qubit basis $\{\ket{\CCC_\alpha^+},\ket{\CCC_\alpha^-}\}$, and the four-photon process, with the logical qubit basis $\{\ket{\CCC_\alpha^{(0\textrm{\scriptsize mod}4)}},\ket{\CCC_{\alpha}^{(2\textrm{\scriptsize mod}4)}}\}$. \textbf{(a)} Considering the two-photon process and initializing the effective two-qubit system in the state $\ket{+_Z,+_Z}=\ket{\CCC_\alpha^+}\otimes\ket{\CCC_\alpha^+}$, we monitor continuously the fidelity with respect to the Bell states $\ket{\BBB_{2,\alpha}^{\pm}}=\frac{1}{\sqrt 2}(\ket{\CCC_\alpha^+}\otimes\ket{\CCC_\alpha^+}\pm i \ket{\CCC_\alpha^-}\otimes\ket{\CCC_\alpha^-})$. The simulation parameters are the same as in previous figures and the effective entangling Hamiltonian is given by $\Omega_{XX}\sigma_x^{1,L}\otimes \sigma_x^{2,L}$ with $\Omega_{XX}=2\bar n \epsilon_{XX}$ ($\epsilon_{XX}=\kappa_{2\textrm{\scriptsize ph}}/20$). \textbf{(b)} Similar simulation for the four-photon process, where the effective two-qubit system is initialized in the state $\ket{+_Z,+_Z}=\ket{\CCC_\alpha^{(0\textrm{\scriptsize mod}4)}}\otimes\ket{\CCC_\alpha^{(0\textrm{\scriptsize mod}4)}}$  and we monitor continuously the fidelity with respect to the Bell states $\ket{\BBB_{4,\alpha}^{\pm}}=\frac{1}{\sqrt 2}(\ket{\CCC_\alpha^{(0\textrm{\scriptsize mod}4)}}\otimes\ket{\CCC_\alpha^{(0\textrm{\scriptsize mod}4)}}\pm i \ket{\CCC_\alpha^{(2\textrm{\scriptsize mod}4)}}\otimes\ket{\CCC_\alpha^{(2\textrm{\scriptsize mod}4)}})$. }}
\label{fig:ZZgate}
\end{figure}

Here we show that the same kind of idea can be applied to the case of two logical qubits to produce an effective entangling Hamiltonian of the form $\sigma_x^L\otimes \sigma_x^L$. We  start with the case of two harmonic oscillators (with corresponding field mode operators $\aaa_1$ and $\aaa_2$), each one undergoing a two-photon  process. Let us assume we can effectively couple these two oscillators to achieve a beam-splitter Hamiltonian of the form $\epsilon_{XX}(\aaa_1\aaa_2^\dag+\aaa_2\aaa_1^\dag)$, where $\epsilon_{XX}\ll\kappa_{1, 2\textrm{\scriptsize ph}}, \kappa_{2, 2\textrm{\scriptsize ph}}$ (we will present in the next section an architecture allowing to get such an effective beam-splitter Hamiltonian between two modes). In order to illustrate the performance of the method, we simulate the two-mode master equation:
\begin{eqnarray}\label{eq:ZZ_2Cat}
\fl \td{\rho}=-i\epsilon_{XX}[\aaa_1\aaa_2^\dag+\aaa_2 \aaa_1^\dag,\rho] +\epsilon_{1, 2\textrm{\scriptsize ph}}[\aaa_1^{\dag 2}-\aaa_1^2,\rho]&+\epsilon_{2, 2\textrm{\scriptsize ph}}[\aaa_2^{\dag 2}-\aaa_2^2,\rho]\\
&+\kappa_{1, 2\textrm{\scriptsize ph}}\DDD[\aaa_1^2]\rho+\kappa_{2, 2\textrm{\scriptsize ph}}\DDD[\aaa_2^2]\rho.
\end{eqnarray} 
Simulations in Fig.~\ref{fig:ZZgate}(a) are performed by initializing the system at the logical state $\ket{+_Z, +_Z}=\ket{\CCC_\alpha^+}\otimes\ket{\CCC_\alpha^+}$ and letting it evolve under Eq. (\ref{eq:ZZ_2Cat}). These simulations illustrate that two-mode entanglement does occur, reaching the Bell states $\ket{\BBB_{2,\alpha}^+}=(\ket{\CCC_\alpha^+}\otimes\ket{\CCC_\alpha^+}+i\ket{\CCC_\alpha^-}\otimes\ket{\CCC_\alpha^-})/\sqrt{2}$ and $\ket{\BBB_{2,\alpha}^-}=(\ket{\CCC_\alpha^+}\otimes\ket{\CCC_\alpha^+}-i\ket{\CCC_\alpha^-}\otimes\ket{\CCC_\alpha^-})/\sqrt{2}$. Indeed, by projecting the beam-splitter Hamiltonian $\epsilon_{XX}(\aaa_1\aaa_2^\dag+\aaa_2 \aaa_1^\dag)$ on the tensor product of the spaces spanned by $\{\ket{\CCC_\alpha^+},\ket{\CCC_\alpha^-}\}$, we get as the effective Hamiltonian, that of a two-qubit entangling gate:
\begin{eqnarray*}
&\fl  \epsilon_{XX}\Pi_{\ket{\CCC_\alpha^+},\ket{\CCC_\alpha^-}}\otimes \Pi_{\ket{\CCC_\alpha^+},\ket{\CCC_\alpha^-}} (\aaa_1\aaa_2^\dag+\aaa_2\aaa_1^{\dag})\Pi_{\ket{\CCC_\alpha^+},\ket{\CCC_\alpha^-}}\otimes \Pi_{\ket{\CCC_\alpha^+},\ket{\CCC_\alpha^-}}=\\
& \fl\qquad\qquad 2|\alpha|^2\epsilon_{XX}\left(\ket{\CCC_\alpha^+}\bra{\CCC_\alpha^-}+\ket{\CCC_\alpha^-}\bra{\CCC_\alpha^+}\right)\otimes\left(\ket{\CCC_\alpha^+}\bra{\CCC_\alpha^-}+\ket{\CCC_\alpha^-}\bra{\CCC_\alpha^+}\right)=\Omega_{XX} \sigma_x^{1,L}\otimes\sigma_x^{2,L},
\end{eqnarray*}
where 
$$
\Omega_{XX}=2\bar n\epsilon_{XX}.
$$
Once again the decay of the fidelity to the Bell states is due to higher order terms in the above approximation of the beam-splitter Hamiltonian by the projected one on the qubit's subspace. This decay can be reduced by taking a larger separation of time-scales between $\epsilon_{XX}$ and $\kappa_{1, 2\textrm{\scriptsize ph}}, \kappa_{2, 2\textrm{\scriptsize ph}}$. However, as can be seen in the simulations, even with a moderate ratio $1/20$ of $\epsilon_{XX}/\kappa_{1, 2\textrm{\scriptsize ph}}$ and $\epsilon_{XX}/\kappa_{2, 2\textrm{\scriptsize ph}}$, we get a Bell state with fidelity in excess of $99\%$.  

For the case of the four-photon process, in order to achieve an effective Hamiltonian of the form $\sigma_x^L\otimes\sigma_x^L$ for the logical qubit basis of $\{\ket{\CCC_\alpha^{(0\textrm{\scriptsize mod}4)}},\ket{\CCC_{\alpha}^{(2\textrm{\scriptsize mod}4)}}\}$, one needs to ensure exchanges of photons in pairs between the two  oscillators encoding the information. This is satisfied by replacing the beam-splitter Hamiltonian with $\epsilon_{XX}\left(\aaa_1^2\aaa_2^{\dag 2}+\aaa_2^2\aaa_{1}^{\dag 2}\right)$. Once again, we initialize the system in the state $\ket{+_Z,+_Z}=\ket{\CCC_{\alpha}^{(0\textrm{\scriptsize mod}4)}}\otimes \ket{\CCC_{\alpha}^{(0\textrm{\scriptsize mod}4)}}$ and we let it evolve following the two-mode master equation:
\begin{eqnarray}\label{eq:ZZ_4Cat}
\fl \td{\rho}=-i\epsilon_{XX}[\aaa_1^2\aaa_2^{\dag 2}+\aaa_2^2 \aaa_1^{\dag 2},\rho] +\epsilon_{1, 4\textrm{\scriptsize ph}}[\aaa_1^{\dag 4}-\aaa_1^4,\rho]&+\epsilon_{2, 4\textrm{\scriptsize ph}}[\aaa_2^{\dag 4}-\aaa_2^4,\rho]\\
&+\kappa_{1, 4\textrm{\scriptsize ph}}\DDD[\aaa_1^4]\rho+\kappa_{2, 4\textrm{\scriptsize ph}}\DDD[\aaa_2^4]\rho.
\end{eqnarray} 
Simulations of Fig.~\ref{fig:ZZgate}(b), illustrate the two-mode entanglement reaching the Bell states $\ket{\BBB_{4,\alpha}^+}=(\ket{\CCC_{\alpha}^{(0\textrm{\scriptsize mod}4)}}\otimes\ket{\CCC_{\alpha}^{(0\textrm{\scriptsize mod}4)}}+i\ket{\CCC_{\alpha}^{(2\textrm{\scriptsize mod}4)}}\otimes\ket{\CCC_{\alpha}^{(2\textrm{\scriptsize mod}4)}})/\sqrt{2}$ and $\ket{\BBB_{4,\alpha}^-}=(\ket{\CCC_{\alpha}^{(0\textrm{\scriptsize mod}4)}}\otimes\ket{\CCC_{\alpha}^{(0\textrm{\scriptsize mod}4)}}-i\ket{\CCC_{\alpha}^{(2\textrm{\scriptsize mod}4)}}\otimes\ket{\CCC_{\alpha}^{(2\textrm{\scriptsize mod}4)}})/\sqrt{2}$. By projecting the Hamiltonian $\epsilon_{XX}(\aaa_1^2\aaa_2^{\dag 2}+\aaa_2^2 \aaa_1^{\dag 2})$ on the tensor product of the spaces spanned by $\{\ket{\CCC_\alpha^{(0\textrm{\scriptsize mod}4)}},\ket{\CCC_{\alpha}^{(2\textrm{\scriptsize mod}4)}}\}$, we get as the effective Hamiltonian, that of a two-qubit entangling gate:
\begin{eqnarray*}
&\fl \epsilon_{XX}\Pi_{\ket{\CCC_\alpha^{(0,2\textrm{\scriptsize mod}4)}}}\otimes \Pi_{\ket{\CCC_\alpha^{(0,2\textrm{\scriptsize mod}4)}}} (\aaa_1^2\aaa_2^{\dag 2}+\aaa_2^2\aaa_1^{\dag 2})\Pi_{\ket{\CCC_\alpha^{(0,2\textrm{\scriptsize mod}4)}}}\otimes \Pi_{\ket{\CCC_\alpha^{(0,2\textrm{\scriptsize mod}4)}}}=\\
&\fl \qquad\qquad\quad 2|\alpha|^4\epsilon_{XX}\left(\ket{\CCC_\alpha^{(0\textrm{\scriptsize mod}4)}}\bra{\CCC_\alpha^{(2\textrm{\scriptsize mod}4)}}+\ket{\CCC_{\alpha}^{(2\textrm{\scriptsize mod}4)}}\bra{\CCC_{\alpha}^{(0\textrm{\scriptsize mod}4)}}\right)^{\otimes 2}=\Omega_{XX} \sigma_x^{1,L}\otimes\sigma_x^{2,L},
\end{eqnarray*}
where 
$$
\Omega_{XX}=2\bar n^2\epsilon_{XX}.
$$

\subsection{Kerr effect for $\pi/2$-rotation around $Z$-axis}\label{ssec:kerr}
In order to achieve a complete set of universal gates, we only need another single-qubit gate consisting of  a $\pi/2$-rotation around the $Y$ or $Z$ axis. Together with arbitrary rotations around $X$-axis, such a single-qubit gate enables us to   perform any unitary operations on single qubits and along with the two-qubit entangling gate of the previous subsection, provides a complete set of universal gates. However, this fixed angle single-qubit gate presents an issue not manifested in the other gates. To see this, consider the case of the two-photon process with the logical qubit basis $\{\ket{\CCC_\alpha^+},\ket{\CCC_\alpha^-}\}$. The process renders the two qubit states $\ket{\pm_X}\approx\ket{\pm\alpha}$ highly stable and tends to prevent any transfer of population from the vicinity of one of these states to the other one. This is trivially in contradiction with the aim of the $\pi/2$-rotation around the $Y$ or $Z$ axis. This simple fact suggests that performing such a gate is not possible in presence of the two-photon process. Here, we propose an alternative approach, consisting of turning off the two-photon process during the operation (possible through the scheme proposed in the next section) and applying a self-Kerr Hamiltonian of the form $-\chi_{\textrm{\scriptsize Kerr}}(\aaa^\dag\aaa)^2$. In the next section, we will see how such a Kerr Hamiltonian is naturally produced through the same setting as the one required for the two-photon processes. 

It was proposed in~\cite{Yurke-Stoler-1986} and experimentally realized in~\cite{Kirchmair-al-Nature_2013} that  a Kerr interaction can be used to generate Schr\"odinger cat states. More precisely, initializing the oscillator in the coherent state $\ket{\beta}$, at any time $t_q=\pi/q\chi_{\textrm{\scriptsize Kerr}}$ where $q$ is a positive integer, the state of the oscillator can be written as a superposition of $q$ coherent states~\cite{haroche-raimond:book06}:
$$
\ket{\psi(t_q=\frac{\pi}{q~\chi_{\textrm{\tiny Kerr}}})}=\frac{1}{2q}\sum_{p=0}^{2q-1}\sum_{k=0}^{2q-1}e^{ik(k-p)\frac{\pi}{q}}\ket{\beta e^{i p\frac{\pi}{q}}}.
$$
In particular, at $t_2=\pi/2\chi_{\textrm{\scriptsize Kerr}}$, the states $\ket{\pm \alpha}$ evolve to $1/\sqrt{2}\left(\ket{\pm \alpha}-i\ket{\mp \alpha}\right)$. Therefore, in the case of the logical qubit basis $\{\ket{\CCC_\alpha^+},\ket{\CCC_\alpha^-}\}$, this is equivalent to a $(-\pi/2)$-rotation around the $Z$-axis.

Analogously for the case of four-photon process, initializing the oscillator in the two-component Schr\"odinger cat state $\ket{+_X}\approx\ket{\CCC_{\alpha}^+}$, obtains the state $1/\sqrt{2}\left(\ket{\CCC_{\alpha}^+}-i\ket{\CCC_{i\alpha}^+}\right)$ at time $t_8=\pi/8\chi_{\textrm{\scriptsize Kerr}}$. Thus, we have a $(-\pi/2)$-rotation around the $Z$-axis for the logical qubit basis of $\{\ket{\CCC_\alpha^{(0\textrm{\scriptsize mod}4)}},\ket{\CCC_\alpha^{(2\textrm{\scriptsize mod}4)}}\}$, .

\subsection{Fault-tolerance}{
\begin{table}
\fl\caption{\label{Table:gates}  List of Hamiltonians and decay operators providing protection and a set of universal gates}
\footnotesize\rm
\begin{tabular*}{\textwidth}{@{}l*{15}{@{\extracolsep{0pt plus12pt}}l}}
\br
 & Two-photon protection & Four-photon protection\\
\mr
Decay operator & $\kappa_{2\textrm{\scriptsize ph}}\aaa^2$ & $\kappa_{4\textrm{\scriptsize ph}}\aaa^4$\\
Driving Hamiltonian & $i\epsilon_{2\textrm{\scriptsize ph}}(\aaa^{\dag2}-\aaa^{2})$ & $i\epsilon_{4\textrm{\scriptsize ph}}(\aaa^{\dag4}-\aaa^{4})$\\
Arbitrary rotations around $X$& $\epsilon_X(\ba^\dag+\ba)$ & $\epsilon_X(\ba^{\dag 2}+\ba^2)$\\
$\pi/2$-rotation around $Z$ & $-\chi_{\textrm{\scriptsize Kerr}}(\ba^\dag\ba)^2$ & $-\chi_{\textrm{\scriptsize Kerr}}(\ba^\dag\ba)^2$ \\
Two-qubit entangling gate & $ \epsilon_{XX}(\ba_1\ba_2^\dag+\ba_2\ba_1^\dag)$ & $\epsilon_{XX}(\ba_1^2\ba_2^{\dag 2}+\ba_2^2\ba_1^{\dag 2})$\\
\br
\end{tabular*}
\end{table}}
{The proposed  set of Hamiltonians  allows one to obtain a set of universal quantum gates for the two cases of two-photon and four-photon processes (see Table~\ref{Table:gates}).} In this subsection, we consider a  logical qubit encoded by the four-photon driven dissipative process and protected against single-photon decay  through continuous photon-number parity measurements. We will  discuss the fault-tolerance of the above single and two qubit gates with respect to the decoherence channels of the single-photon decay and the photon dephasing. Indeed, we will not discuss here the tolerance with respect to imprecisions of the gates themselves as we believe such errors should not be put on the same footing as the errors induced by the decoherence of the involved quantum systems. While the protection against errors due to the coupling to an uncontrolled environment is crucial to ensure a scaling towards many-qubit quantum computation, the degree of perfection of gate parameters, such as the angle of a rotation for instance, can be regarded as a technical and engineering matter.  

More precisely, we show that the error rate due to the photon loss channel does not  increase while performing the quantum operations of the previous subsections and that  the continuous parity measurements during the operations enable the protection against such a decay channel. Furthermore, arbitrary rotations of the single-qubit around $X$-axis as well as the two-qubit entangling gate are performed in presence of the four-photon process and therefore the qubit will also remain protected against photon dephasing channel. For the  single-qubit $\pi/2$-rotation around $Z$-axis, as long as the Kerr Hamiltonian strength $\chi_{\textrm{\tiny Kerr}}$ is much more prominent than the dephasing rate (which is the case in most current circuit QED schemes), turning on the four-photon process after the operation will correct for the phase error accumulated during the operation. 

\begin{figure}[h]
\begin{center}
\includegraphics[width=\textwidth]{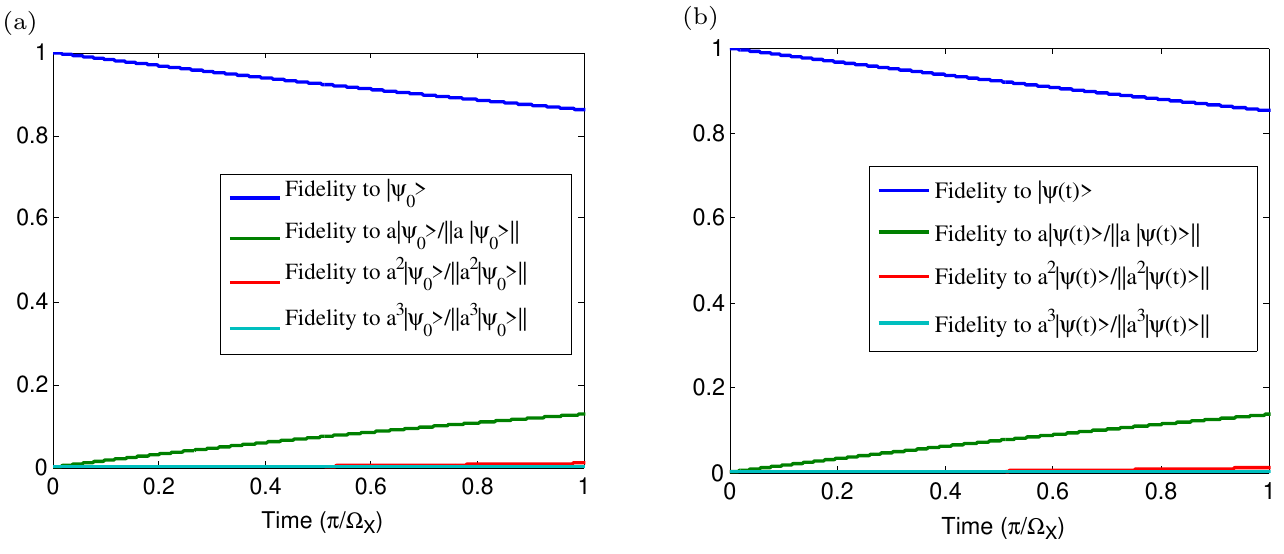}
\end{center}
\caption{Decay of the unprotected qubit (no photon number parity measurements) encoded in the 4-cat scheme due to single-photon loss channel. \textbf{(a)} the qubit is initialized in the state $\ket{\psi_0}=\ket{\CCC_\alpha^{(0\textrm{\scriptsize mod}4)}}$ and no gate is applied on the qubit. The decoherence due to the single-photon loss channel leads to a decay in the fidelity with respect to the initial state and creates a mixture of this state with the three other states $\aaa\ket{\psi_0}/\|\aaa\ket{\psi_0}\|=\ket{\CCC_\alpha^{(3\textrm{\scriptsize mod}4)}}$, $\aaa^2\ket{\psi_0}/\|\aaa^2\ket{\psi_0}\|=\ket{\CCC_\alpha^{(2\textrm{\scriptsize mod}4)}}$ and $\aaa^3\ket{\psi_0}/\|\aaa^3\ket{\psi_0}\|=\ket{\CCC_\alpha^{(1\textrm{\scriptsize mod}4)}}$. \textbf{(b)} in presence of the squeezing Hamiltonian performing the $X_\theta$ operation, this decoherence rate remains similar and mixes the desired state $\ket{\psi(t)}=\cos(\Omega_X t)\ket{\CCC_\alpha^{(0\textrm{\scriptsize mod}4)}}-i\sin(\Omega_X t)\ket{\CCC_\alpha^{(2\textrm{\scriptsize mod}4)}}$  with the states $\aaa\ket{\psi(t)}/\|\aaa\ket{\psi(t)}\|=\cos(\Omega_X t)\ket{\CCC_\alpha^{(3\textrm{\scriptsize mod}4)}}-i\sin(\Omega_X t)\ket{\CCC_\alpha^{(1\textrm{\scriptsize mod}4)}}$, $\aaa^2\ket{\psi(t)}/\|\aaa^2\ket{\psi(t)}\|=\cos(\Omega_X t)\ket{\CCC_\alpha^{(2\textrm{\scriptsize mod}4)}}-i\sin(\Omega_X t)\ket{\CCC_\alpha^{(0\textrm{\scriptsize mod}4)}}$ and $\aaa^3\ket{\psi(t)}/\|\aaa^3\ket{\psi(t)}\|=\cos(\Omega_X t)\ket{\CCC_\alpha^{(1\textrm{\scriptsize mod}4)}}-i\sin(\Omega_X t)\ket{\CCC_\alpha^{(3\textrm{\scriptsize mod}4)}}$. The photon jumps inducing such mixing of the quantum states are however tractable through continuous photon number parity measurements.}
\label{fig:tolerance}
\end{figure}

\textit{Single-qubit $X_\theta$ gate and two-qubit entangling gate.}  These operations would be performed in concurrence with the  four-photon process, which continuously and strongly projects to the state space generated by $\{\ket{\pm\alpha},\ket{\pm i\alpha}\}$. Consider the case of single-qubit $X_\theta$ gate (with the same kind of analysis  valid for the two-qubit entangling gate). Starting with the state $\ket{+_Z}=\ket{\CCC_\alpha^{(0\textrm{\scriptsize mod}4)}}$ and in the absence of single-photon jumps, the system evolves at time $t$ to $\ket{\psi(t)}=\cos(\Omega_X t)\ket{\CCC_\alpha^{(0\textrm{\scriptsize mod}4)}}-i\sin(\Omega_X t)\ket{\CCC_\alpha^{(2\textrm{\scriptsize mod}4)}}$. With the additional presence of one single-photon jump during this time, this state becomes $\aaa\ket{\psi(t)}=\cos(\Omega_X t)\ket{\CCC_\alpha^{(3\textrm{\scriptsize mod}4)}}-i\sin(\Omega_X t)\ket{\CCC_\alpha^{(1\textrm{\scriptsize mod}4)}}$. More precisely, after a single-photon jump has occurred, the Zeno dynamics of Eq.~(\ref{eq:Zeno4Cat}) keeps ensuring the rotation around the $X$-axis of the new logical qubit basis $\{\ket{\CCC_\alpha^{(3\textrm{\scriptsize mod}4)}},\ket{\CCC_\alpha^{(1\textrm{\scriptsize mod}4)}}\}$ corresponding to the odd photon number parity manifold. Note that after two and three photon jumps, we respectively get back to the even and odd parity manifolds but  altering the basis elements (equivalent to a bit-flip). Finally, after four jumps, we end up in the initial logical basis as if no jump has occurred. This simple reasoning indicates that a continuous photon number parity measurement during the operation should ensure the protection of the rotating quantum information against the single-photon decay channel. The simulations of Fig.~\ref{fig:tolerance} confirm the fact that performing such a single qubit $X_\theta$ gate, in the presence of the single-photon decay channel, does not increase the decay rate or lead to new decay channels. Continuous photon number parity measurements should therefore correct for such loss events and protect the qubit while the operation is performed.  These simulations correspond to the master equation:
\begin{eqnarray*}
\td\rho=-i\epsilon_X[\aaa^2+\aaa^{\dag 2},\rho]+\epsilon_{4\textrm{\scriptsize ph}}[\aaa^{\dag 4}-\aaa^4,\rho]+\kappa_{4\textrm{\scriptsize ph}}\DDD[\aaa^4]\rho+\kappa_{1\textrm{\scriptsize ph}}\DDD[\aaa]\rho.
\end{eqnarray*}
We take $\epsilon_X=0$ and $\epsilon_X=\kappa_{4\textrm{\scriptsize ph}}/20$ respectively in Figures~\ref{fig:tolerance}(a) and (b) and $\kappa_{1\textrm{\scriptsize ph}}=\kappa_{4\textrm{\scriptsize ph}}/200$ for both plots. As can be seen through these plots, the decay rate remains the same in absence or presence of the two-photon driving field ensuring the arbitrary rotation around the $X$-axis. Additionaly, the probability of having more than one jump during the operation time remains within the range of $1\%$, indicating that with such parameters one would not even  need to perform photon-number  parity measurements during the operation and that a measurement after the operation would be enough to ensure a significant improvement in the coherence time.

\textit{Single-qubit $\pi/2$-rotation around $Z$-axis.} In order to show that the Kerr effect can be applied in a fault-tolerant manner to perform such a single-qubit operation, we apply some of the arguments of the supplemental material of~\cite{Leghtas-al-PRL-2013}. We need to consider the effect of photon loss events on the logical qubit  during such an operation. 

We note first that the unitary generated by the Kerr Hamiltonian does not modify the photon number parity as this Hamiltonian is diagonal in the Fock states basis.  Therefore, photon number parity remains a quantum jump indicator in presence of the Kerr effect. Now, let us assume that a jump occurs at time $t$ during the operation: the state after the jump is given by 
$$
\aaa e^{it \chi_{\tiny Kerr}(\aaa^\dag\aaa)^2}\ket{\psi_0}=e^{2it \chi_{\tiny Kerr}\aaa^\dag\aaa}e^{it \chi_{\tiny Kerr}(\aaa^\dag\aaa)^2} \aaa \ket{\psi_0},
$$
where we have applied the commutation relation $\aaa f(\aaa^\dag\aaa)=f(\aaa^\dag\aaa+\II)\aaa$, $f$ being an arbitrary analytic function. This means that up to a phase space rotation $e^{i2t \chi_{\tiny Kerr}\aaa^\dag\aaa}$, the effect of a photon jump event commutes with the unitary generated by the Kerr Hamiltonian. Assuming much faster parity measurements than the Kerr dynamics and keeping track of both the number of parity jumps $p$ and the times of their occurrences $\{t_k\}_{k=1}^p$, the state after the operation is fully known. In particular, the four-component Schr\"odinger cat state is rotated in  phase space by an angle of $2(\sum_{k=1}^p t_k)\chi_{\scriptsize Kerr}$. We can take this phase space rotation into account by merely changing the phase of the four-photon drive $\epsilon_{4\textrm{\scriptsize ph}}$ in the four-photon process.  

\section{Towards an experimental realization within a circuit QED framework}
\label{sec:realization}

\subsection{Two-photon driven dissipative process}
\label{subsec:twophpump}
\begin{figure}[htb!] 
\centering
\includegraphics[width=9cm]{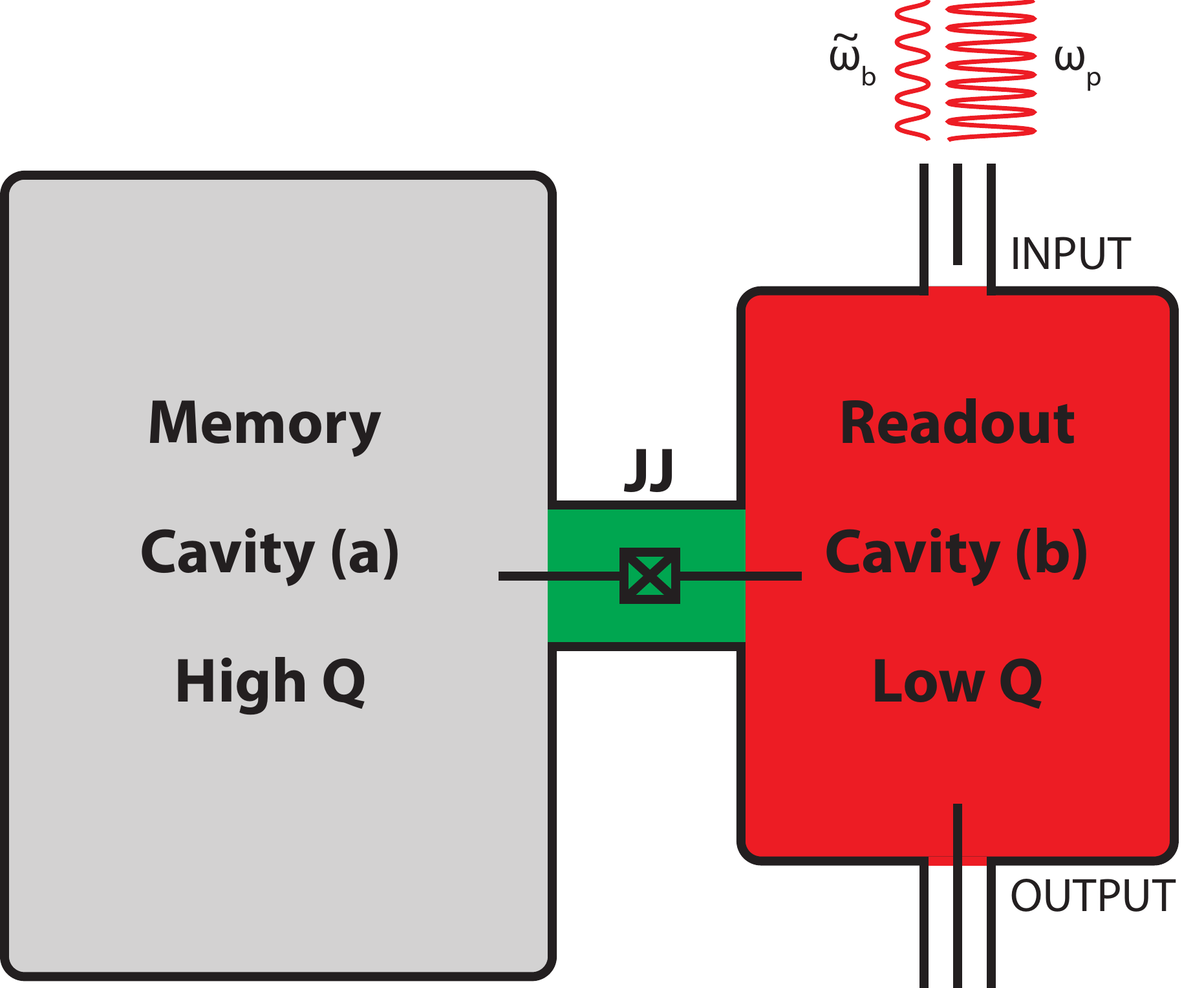}
\caption{ Proposal for a practical realization of the two photon driven dissipative process. Two cavities are linked by a small transmission line in which a Josephson tunnel junction is embedded. This element provides a non-linear coupling between the modes of these two cavities. A pump tone at frequency $\omega_p$ is applied to the readout cavity. If we set $\omega_p=2\tilde\omega_a-\tilde\omega_b$ ($\tilde\omega_a$ and $\tilde\omega_b$ are the shifted frequencies of the modes $\ba$ and $\bb$ in the presence of all couplings and the pump), we select an interaction term of the form $\ba^2\bb^\dag + \textrm{c.c.}$, where $\ba$ and $\bb$ are the annihilation operators for the fundamental modes of the cavities. Combining this interaction with a drive and strong single-photon dissipation of mode $\bb$ leads to the desired dynamics for mode $\ba$ of the form Eq.~(\ref{eq:Two-Photon}). In this way, quantum information can be stored and protected in mode $\ba$.}
\label{fig:TwoCavities}
\end{figure}

In this subsection, we propose an architecture based on Josephson circuits which implements the two photon driven dissipative process. Using the coupling of cavity modes to a Josephson junction (JJ), single photon dissipation, and coherent drives, we aim to produce  effective dynamics in the form of Eq.~(\ref{eq:Two-Photon}). These are the same tools used in the Josephson Bifurcation Amplifier (JBA) to produce a squeezing Hamiltonian~\cite{Vijay-et-al-09} and here we will show that, by selecting a particular pump  frequency, we can achieve a two photon driven dissipative process. Furthermore, in the next subsection, we show that by choosing adequate pump frequencies, we may engineer the interaction terms needed to perform the logical gates described in subsections \ref{sec:zenophaseshiftgate},~\ref{sec:zenotwoqubitentanglement} and~\ref{ssec:kerr}. An architecture suitable for the four photon driven dissipative process is subject to ongoing work.

The practical device we are considering is represented in Fig.~\ref{fig:TwoCavities}. Two cavities are linked by a small transmission line in which a Josephson Junction is embedded. This provides a non-linear coupling between the modes of these two cavities~\cite{Kirchmair-al-Nature_2013,Vlastakis-et-al-Science_2013}. The Hamiltonian of this device is given by \cite{nigg-et-al-2012} 
\begin{eqnarray}
\label{eq:Hnot}
\bH_0=\sum_k{\hbar\omega_k \ba_k^\dag \ba_k}-E_J\cos\left(\frac{\bPhi}{\phi_0}\right),\qquad \bPhi &=& \sum_k{\phi_k(\ba_k+\ba_k^\dag)}\;,
\end{eqnarray}
where $E_J$ is the Josephson energy, $\phi_0=\hbar/2e$ is the reduced superconducting flux quantum, and $\phi_p$ is the zero point flux fluctuation for mode $p$ of frequency $\omega_p$.
Here we are only concerned by the dynamics of the fundamental modes of the two cavities and we assume that all  other modes are never excited. We denote $\ba$ and $\bb$ the annihilation operators of these two modes and $\omega_a$, $\omega_b$ their respective frequencies. We assume that $\mid\bPhi/\phi_0\mid \ll 1$ so that we can neglect sixth and higher order terms in the expansion of the cosine. In order to select the terms of interest, we propose to drive mode $\bb$ with two fields: a weak resonant drive $\epsilon_b(t)$ and a strong off-resonant pump $\epsilon_p(t)$.
The frequencies of modes $\ba$ and $\bb$ are shifted by the non-linear coupling. The dressed frequencies are noted $\tilde\omega_a$ and $\tilde\omega_b$ and we take $\epsilon_b(t)=2\epsilon_b \cos(\tilde\omega_b t)$ and $\epsilon_p(t) = 2\epsilon_p \cos(\omega_p t)$ with :
$$
\omega_p = 2 \tilde\omega_a - \tilde\omega_b.
$$
We place ourselves in a regime where rotating terms can be neglected and the remaining terms after the rotating wave approximation constitute the effective Hamiltonian

\begin{equation}
\label{eq:HRWA2ph}
\fl \frac{1}{\hbar}\overline\bH_{2\textrm{\scriptsize ph}}=g_{2\textrm{\scriptsize ph}}(\ba^2 \bb^\dag + \ba^{\dag 2} \bb) -\epsilon_b(\bb^\dag+\bb) + \frac{\chi_{aa}}{2}(\ba^\dag \ba)^2+\frac{\chi_{bb}}{2}(\bb^\dag \bb)^2 +\chi_{ab}(\ba^\dag \ba)(\bb^\dag \bb) \;.
\end{equation}
While the induced self-Kerr and cross-Kerr terms $\chi_{aa}$, $\chi_{bb}$ and $\chi_{ab}$ can be deduced  from the Hamiltonian of Eq.~(\ref{eq:Hnot}) through the calculations of~\cite{nigg-et-al-2012}, one  similarly finds 
$$
g_{2\textrm{\scriptsize ph}}=\frac{\epsilon_p}{\omega_p-\tilde\omega_b}\chi_{ab}/2.
$$
More precisely, this model reduction can be done by going to a displaced rotating frame in which the Hamiltonian of the pumping drive is removed. Next, one develops the cosine term in the Hamiltonian of~Eq.~(\ref{eq:Hnot}) up to the fourth order and removes the highly oscillating terms in a rotating wave approximation. 

Physically, the pump tone $\epsilon_p$ allows two photons of mode $\aaa$ to convert to a single photon of mode $\bb$, which can decay through the lossy channel coupled to mode $\bb$. The drive tone $\epsilon_b$ inputs energy into mode $\bb$, which can then be converted to pairs of photon in mode $\aaa$. The last three terms in Eq.~(\ref{eq:HRWA2ph}) are the Kerr and cross-Kerr couplings inherited from our proposed architecture. Although these are parasitic terms, we show through numerical simulations that their presence does not deteriorate our scheme.

Taking into account single-photon decay of the mode $\bb$, the effective master equation is given by:
\begin{equation}
\label{eq:LindbladRWA}
\td\rho_{2\textrm{\scriptsize ph}} = -\frac{i}{\hbar}[\overline\bH_{2\textrm{\scriptsize ph}},\rho_{2\textrm{\scriptsize ph}}]+\kappa_b\DDD[\bb]\rho_{2\textrm{\scriptsize ph}}\;.
\end{equation}
Neglecting the Kerr and cross Kerr terms and assuming that $\textrm{$g_{2\textrm{\scriptsize ph}}$},\epsilon_b\ll \kappa_b$, we adiabatically eliminate mode $\bb$~\cite{Drummond-et-al-1981,Carmichael-Wolinsky-1988} and find a reduced dynamics for mode $\aaa$ of the form of Eq.~(\ref{eq:Two-Photon}) where 
\begin{equation*}
\epsilon_{2\textrm{\scriptsize ph}}=\frac{2\epsilon_b\textrm{$g_{2\textrm{\scriptsize ph}}$}}{\kappa_b},\; \kappa_{2\textrm{\scriptsize ph}}= \frac{4 g_{2\textrm{\scriptsize ph}}^2}{\kappa_b} \textrm{ and } \alpha=\sqrt{\epsilon_b/g_{2\textrm{\scriptsize ph}}}\;.
\end{equation*}

\begin{figure}[h] 
\centering
\includegraphics[width=9cm]{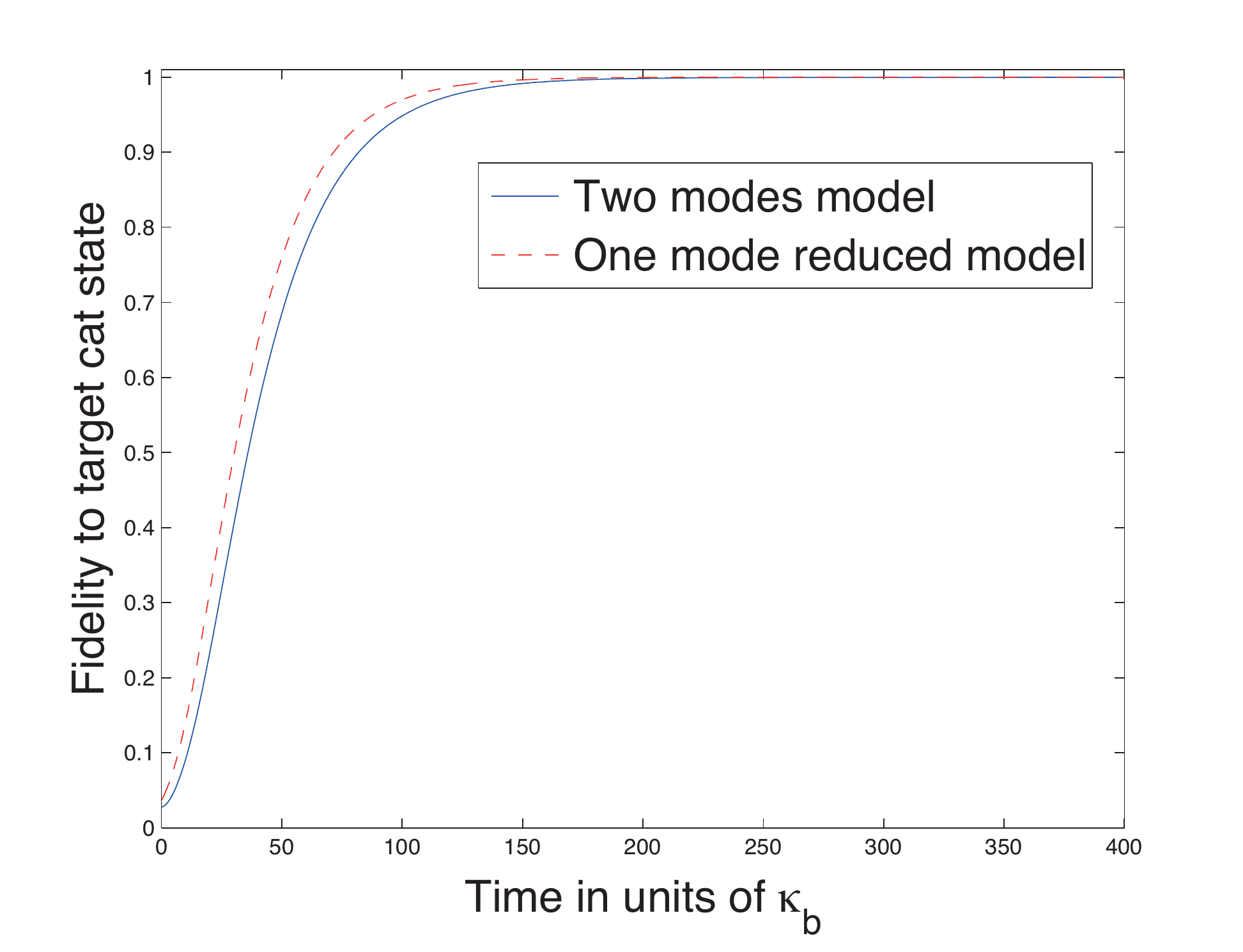}
\caption{Numerical simulation of  Eq.~(\ref{eq:LindbladRWA}) (full blue line) and Eq.~(\ref{eq:Two-Photon})  (dashed red line). We represent the fidelity of the state w.r.t the state $\ket{\CCC_\alpha^+}$, where $\mathcal{N}$ is a normalization factor, and $\alpha=\sqrt{\epsilon_b/g_{2\textrm{\scriptsize ph}}}$. The dashed and full curves have comparable convergence rates and converge to the same state. This indicates that the reduced model of  Eq.~(\ref{eq:Two-Photon})  is a faithful representation of the complete model  Eq.~(\ref{eq:LindbladRWA}). The finite discrepancy is due to the finite ratio between $g_{2\textrm{\scriptsize ph}}, \epsilon_b$ and $\kappa_b$, and the presence of non zero Kerr and cross Kerr terms.}
\label{fidelity2ph}
\end{figure}
One can check the validity of this model reduction by comparing the numerical simulation of Eq.~(\ref{eq:Two-Photon}) to the master equation Eq.~(\ref{eq:LindbladRWA}). Fixing $\kappa_b = 1$\footnote[7]{We have intentionally avoided to provide the units to only focus on the separation of time-scales; however, all these parameters could be considered in the units of $2\pi\times$MHz and  they will be within the reach of current circuit QED setups.} we take $\chi_{aa}=0.0015,\chi_{bb}=0.185,\chi_{ab}=0.033$ and $\epsilon_p/(\tilde\omega_b-\omega_p)=3$, and hence $g_{2\textrm{\scriptsize ph}}=0.05$, $\epsilon_b=4g_{2\textrm{\scriptsize ph}}\textrm{ (to fix the average number of photons in the target cat to 4)}$. In Figure \ref{fidelity2ph}, we compare the fidelity to the target cat state of solutions of Eq.~(\ref{eq:LindbladRWA}) (blue solid line) and solutions of Eq.~(\ref{eq:Two-Photon}), starting in vacuum. The two curves both converge to a fidelity close to one, which indicates that the steady state of Eq.~(\ref{eq:LindbladRWA}) is hardly affected by the presence of Kerr and cross Kerr terms and by the finite ratio of $g_{2\textrm{\scriptsize ph}},\epsilon_b$ to $\kappa_b$.

\subsection{Logical operations}
\begin{figure}[h] 
\centering
\includegraphics[width=\columnwidth]{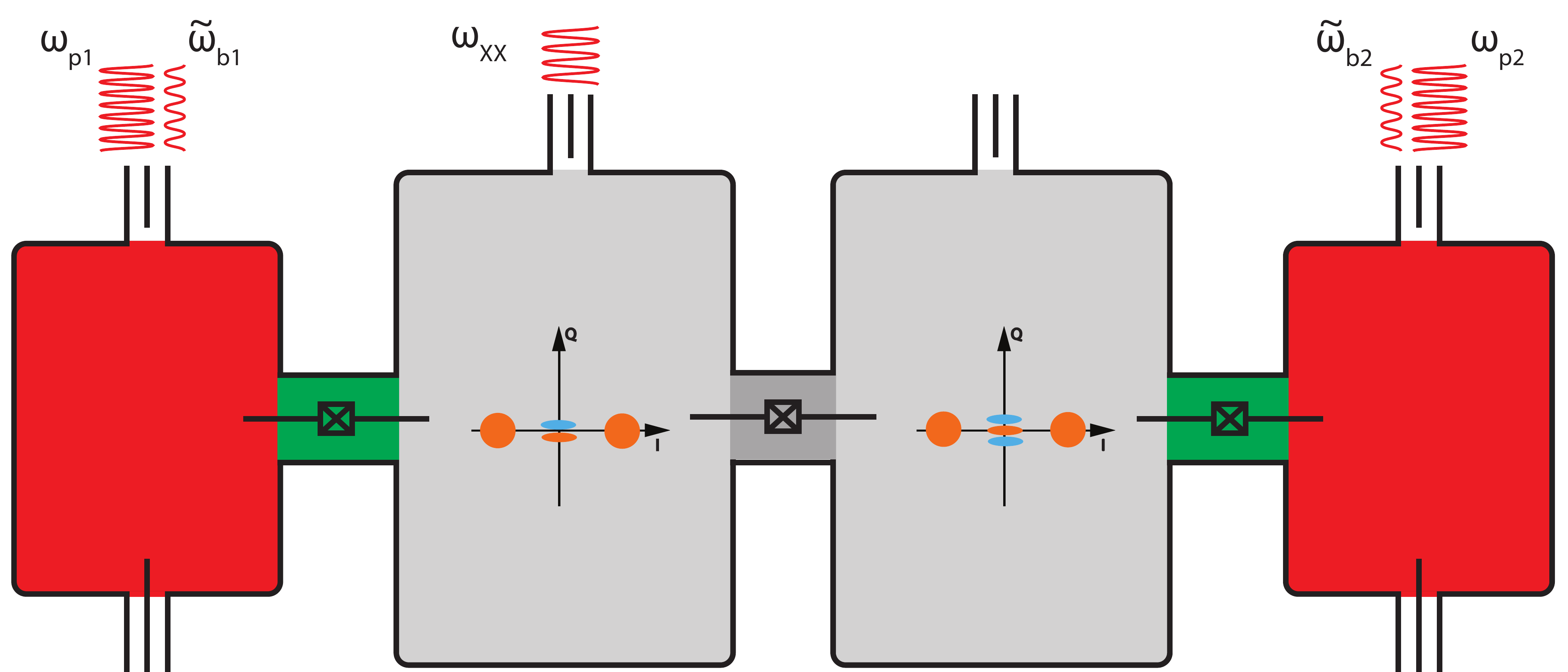}
\caption{Architecture for coupling two qubits protected by the two-photon driven dissipative processes. Two modules composed of a pair of high and low cavities are connected through a central JJ. This JJ provides a nonlinear coupling between the two storage modes $a_1$ and $a_2$ of each module. Adding a pump at frequency $\omega_{XX}=(\tilde\omega_{a1}-\tilde\omega_{a2})/2$ induces an interaction term of the form $\ba_1\ba_2^\dag+c.c$, thus allowing for the entangling gate detailed in Sec.~\ref{sec:zenotwoqubitentanglement}}
\label{fig:TwoModules}
\end{figure}
\paragraph{Rotations of arbitrary angles around the $X$-axis for the logical qubit $\{\ket{\CCC_\alpha^+},\ket{\CCC_\alpha^-}\}$:} simply adding a drive of amplitude $\epsilon_a$ resonant with mode $a$ will add a term proportional to $\epsilon_a^*\ba+\epsilon_a\ba^\dag$ in Eq.~(\ref{eq:Two-Photon}). In the limit where $\abs{\epsilon_a}\ll \kappa_{2ph}$, this will induce coherent oscillation between the two states around the Bloch sphere's $X$-axis, as explained in Sec.~\ref{sec:zenophaseshiftgate}.
\paragraph{Entangling gate between two logical bits:} we propose the architecture of Fig.~\ref{fig:TwoModules} to couple two qubits protected by a two photon driven dissipative process. Two modules, each composed of a pair of high and low Q cavities, are coupled through a JJ embedded in a waveguide connecting the two high Q cavities. This JJ provides a nonlinear coupling, which, together with a pump at frequency $\omega_{ZZ}=(\tilde\omega_{a_1}-\tilde\omega_{a_2})/2$, induces  an interaction of the form $e^{i\phi_{\textrm{\scriptsize pump}}}\ba_1\ba_2^\dag+c.c$. Such a term performs an entangling gate between two logical qubits, as described in Sec.~\ref{sec:zenotwoqubitentanglement}.
\paragraph{$\pi/2$-rotation around $Z$-axis:} as mentioned throughout the previous subsection, the mere fact of coupling the cavity mode to a JJ induces a self-Kerr term on the cavity mode. As proposed in Sec.~\ref{ssec:kerr}, this could be employed to perform a $\pi/2$-rotation around the $Z$-axis in a similar manner to~\cite{Kirchmair-al-Nature_2013}. One only needs to turn off all the pumping drives and wait for $\pi/\chi_{aa}$.

\subsection{{Extension to four-photon driven dissipative process}}\label{ssec:realization-4cat}

\begin{figure}[h]
\includegraphics[width=\textwidth]{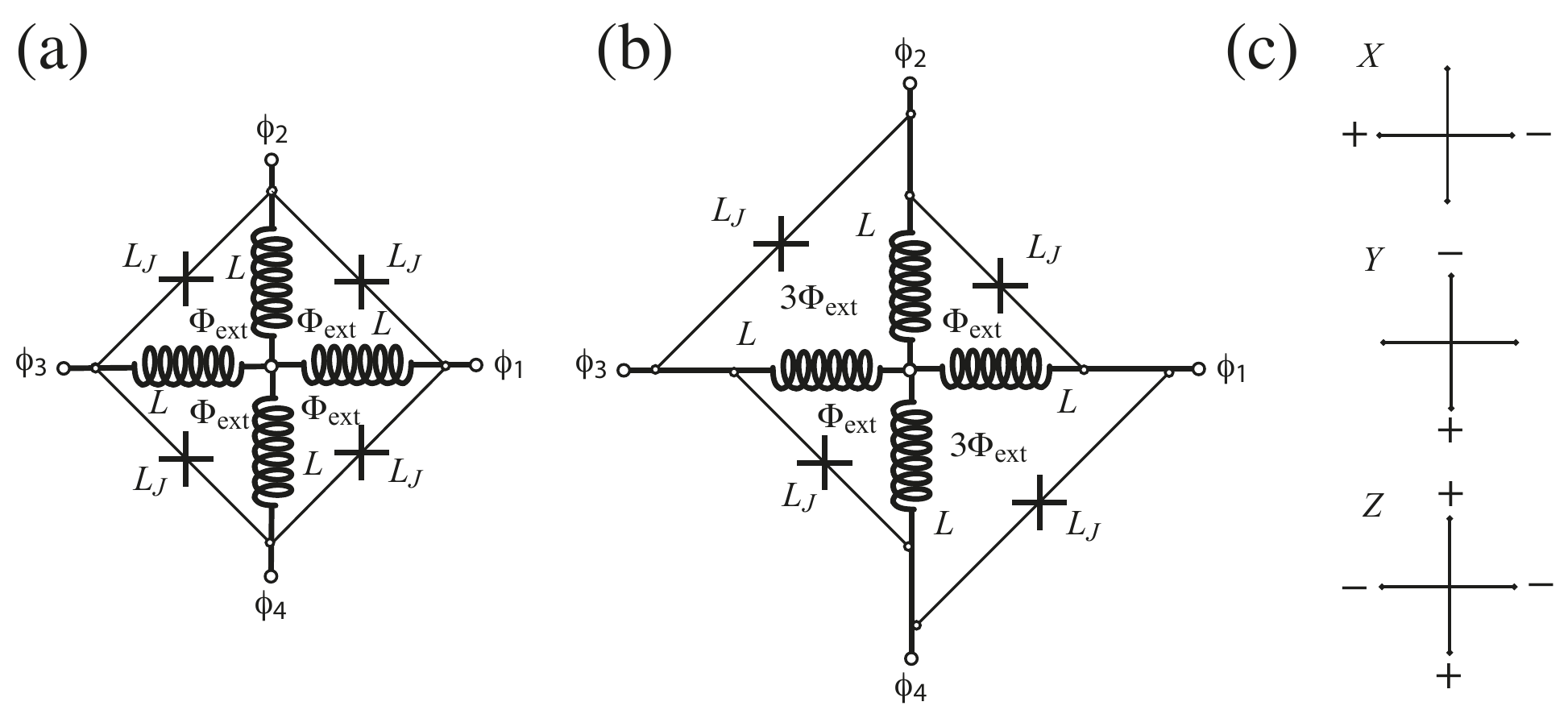}
\caption{Josephson ring modulators (JRM) providing desired interactions between field modes. \textbf{(a)} JRM developed to ensure quantum limited amplification of a quantum signal or to provide frequency conversion between two modes; The signal and idler are respectively coupled to the $X$  and $Y$ modes, as represented in \textbf{(c)} and the pump drive is applied on the $Z$ mode. \textbf{(b)} A modification of the JRM to ensure an interaction of the form Eq.~(\ref{eq:JJ4pump}). Such an interaction should allow us to achieve the driven dissipative four-photon process without adding  undesired Hamiltonian terms.}\label{fig:JRM}
\end{figure}

Similarly to the case of the two-photon process, we need to achieve an effective Hamiltonian of the form
$$
\frac{1}{\hbar}\overline H_{4\textrm{\scriptsize ph}}=g_{4\textrm{\scriptsize ph}}(\ba^4\bb^\dag+\ba^{\dag 4}\bb)+\epsilon_b(\bb+\bb^\dag).
$$
Taking into account the single-photon decay of the mode $\bb$ of rate $\kappa_b$ such that $g_{4\textrm{\scriptsize ph}},\epsilon_b\ll\kappa_b$, we can adiabatically eliminate the mode $\bb$ and find a reduced dynamics for mode $\ba$ of the form Eq.~(\ref{eq:Four-Photon}). The problem is therefore to engineer in an efficient manner the Hamiltonian $\overline H_{4\textrm{\scriptsize ph}}$. 

Indeed, the same architecture as in Fig.~\ref{fig:TwoCavities} together with a pump frequency of $\omega_{p}=4\tilde\omega_a-\tilde\omega_b$ should induce an effective Hamiltonian term of the form $g_{4\textrm{\scriptsize ph}}(\ba^4\bb^\dag+\ba^{\dag 4}\bb)$. One can easily observe this by developing the cosine term in Eq.~(\ref{eq:Hnot}) up to the sixth order in $\bPhi/\phi_0$ and by applying a rotating wave approximation, leading to an effective coupling strength of $g_{4\textrm{\scriptsize ph}}=\frac{E_J}{24\hbar}\frac{\epsilon_p}{\omega_p-\tilde\omega_b}\frac{\phi_a^4\phi_b^2}{\phi_0^6}$. However, such an architecture also leads to other significant terms limiting the performance of the process. In particular, through the same sixth order expansion, one can observe an amplified induced Kerr effect on the mode $\ba$: $\chi_{aa}^{\textrm{\scriptsize pumped}}(\ba^\dag\ba)^2$ with $\chi_{aa}^{\textrm{\scriptsize pumped}}=\frac{E_J}{4\hbar}\frac{\epsilon_p^2}{(\omega_p-\tilde\omega_b)^2}\frac{\phi_a^4\phi_b^2}{\phi_0^6}=\frac{6\epsilon_p}{\omega_p-\tilde\omega_b}g_{4\textrm{\scriptsize ph}}$.

Inspired by the architecture of the Josephson ring modulator~\cite{Bergeal-et-al-2010,Roch-et-al-12}, which ensures an efficient three-wave mixing, we propose here a design which should induce very efficiently the above effective Hamiltonian while avoiding the addition of extra undesirable interactions.  The Josephson ring modulator (Fig.~\ref{fig:JRM}(a)) provides a coupling between the three modes (as presented in Fig.~\ref{fig:JRM}(c)) of the form 
\begin{eqnarray*}
\fl H_{\textrm{\scriptsize JRM}}=\frac{E_L}{4}(\frac{\bPhi_X^2}{\phi_0^2}+\frac{\bPhi_Y^2}{\phi_0^2}+\frac{\bPhi_Z^2}{2\phi_0^2})\\
-4E_J\left[\cos \frac{\bPhi_X}{2\phi_0} \cos \frac{\bPhi_Y}{2\phi_0} \cos \frac{\bPhi_Z}{2\phi_0} \cos \frac{\Phi_{\textrm{\scriptsize ext}}}{\phi_0}+\sin \frac{\bPhi_X}{2\phi_0} \sin \frac{\bPhi_Y}{2\phi_0} \sin\frac{\bPhi_Z}{2\phi_0}\sin\frac{\Phi_{\textrm{\scriptsize ext}}}{\phi_0}\right],
\end{eqnarray*}
where $E_L=\phi_0^2/L$, $\bPhi_{X,Y,Z}=\phi_{X,Y,Z}(\ba_{X,Y,Z}+\ba_{X,Y,Z}^\dag)$ and $\Phi_{\textrm{\scriptsize ext}}/\phi_0$ is the dimensionless external flux threading each of the identical four loops of the device. Furthermore, the three spatial mode amplitudes $\bPhi_X=\bphi_3-\bphi_1$, $\bPhi_Y=\bphi_4-\bphi_2$ and $\bPhi_Z=\bphi_2+\bphi_4-\bphi_1-\bphi_3$ are gauge invariant orthogonal linear combinations of the superconducting phases of the four nodes of the ring (Fig.~\ref{fig:JRM}(c)). 

In the same manner the design of Fig.~\ref{fig:JRM}(b), for a dimensionless external flux of  $\Phi_{\textrm{\scriptsize ext}}/\phi_0=\pi/4$ on the small loops and $3\Phi_{\textrm{\scriptsize ext}}/\phi_0=3\pi/4$ on the big loops, induces an effective interaction Hamiltonian of the form
\begin{eqnarray}\label{eq:JJ4pump}
\fl\qquad H'_{\textrm{\scriptsize JRM}}=\frac{E_L}{4}(\frac{\bPhi_X^2}{\phi_0^2}+\frac{\bPhi_Y^2}{\phi_0^2}+\frac{\bPhi_Z^2}{2\phi_0^2})-2\sqrt{2}E_J\sin\frac{\bPhi_X}{2\phi_0} \sin\frac{\bPhi_Y}{2\phi_0} \left[\sin \frac{\bPhi_Z}{2\phi_0}+ \cos \frac{\bPhi_Z}{2\phi_0}\right].
\end{eqnarray}
Similarly to~\cite{Roch-et-al-12}, by decreasing the inductances $L$ and therefore increasing the associated $E_L$, one can keep the three modes of the device stable for such a choice of external fluxes. This however comes at the expense of diluting the nonlinearity.

Now, we couple the $Z$ mode of the device to the high-Q storage mode $\aaa$, its $Y$ mode to the low-Q $\bb$ mode,  and we drive the $X$ mode by a pump of frequency $4\tilde\omega_a-\tilde\omega_b$ ($\tilde\omega_a$ and $\tilde\omega_b$ are the effective frequencies of the modes $\ba$ and $\bb$). By expanding the Hamiltonian of Eq.~(\ref{eq:JJ4pump}) up to sixth order terms in $\phi=(\frac{\bPhi_X}{\phi_0},\frac{\bPhi_Y}{\phi_0},\frac{\bPhi_Z}{\phi_0})$, the only non-rotating term will be of the form
$$
H_{\textrm{\scriptsize eff}}=-\frac{\sqrt{2}}{768}\sqrt{n_{\textrm{\scriptsize pump}}}E_J\frac{\phi_Z^4\phi_Y\phi_X}{\phi_0^6}(e^{i\phi_{\textrm{\scriptsize pump}}}\ba^4\bb^\dag+e^{-i\phi_{\textrm{\scriptsize pump}}}\ba^{\dag 4}\bb),
$$
where $\phi_{\textrm{\scriptsize pump}}$ is the phase of the pump drive and $n_{\textrm{\scriptsize pump}}$ is the average photon number of the coherent state produced in the pump resonator~\cite{Schackert-thesis}.

\ack

This research was supported by the Intelligence Advanced Research Projects Activity (IARPA) W911NF-09-1-0369 and by the U.S. Army Research Office W911NF-09-1-0514. MM  acknowledges support from the Agence National de Recherche under the project EPOQ2 ANR-09-JCJC-0070. ZL acknowledges support from the NSF DMR 1004406. VVA acknowledges support from the NSF Graduate Research Fellowships Program. LJ acknowledges support from  the Alfred P. Sloan Foundation, the Packard Foundation, and the DARPA Quiness program.
\appendix
\section{Asymptotic behavior of the two- and four-photon processes}\label{append:victor}

\subsection{Asymptotic state for arbitrary initial state of the two-photon process}

As stated in Sec.~\ref{ssec:twoPh}, all initial states evolving under the two-photon driven dissipative process from Eq.~(\ref{eq:Two-Photon}) will exponentially converge to a specific (possibly mixed) asymptotic density matrix defined on the Hilbert space spanned by the two-component Schr\"odinger cat states $\{\cp,\cm\}$ with $\a=\moda e^{i\t_\a}$. In order to characterize the Bloch vector of this asymptotic density matrix $\rinf$ [Eq.~(\ref{eq:rinf})], it is sufficient to determine three degrees of freedom: the population of one of the cats ($c_{++}=\cpb\rinf\cp$) and the complex coherence between the two ($c_{+-}=\cmb\rinf\cp$). There exist conserved quantities $J_{++},J_{+-}$ corresponding to these degrees of freedom~\cite{Albert-Jiang-2013} such that $c_{++}=\tr{J_{++}^\dag \rin}$ and $c_{+-}=\tr{J_{+-}^\dag \rin}$ for any initial state $\rin$. These conserved quantities are given by
\vba
J_{++}=\sum_{n=0}^{\infty}|2n\rangle \langle 2n| \label{eq:j00}\\\label{eq:j01}
J_{+-}=\sqrt{\frac{2\modas}{\sinh\left(2\modas\right)}}\sum_{q=-\infty}^{\infty}\frac{(-1)^{q}}{2q+1}I_{q}(\modas)J_{+-}^{\left(q\right)}e^{-i\t_\a(2q+1)}
,\vea
where $I_q(.)$ is the modified Bessel function of the first kind and
\begin{equation*}
J_{+-}^{\left(q\right)}=\begin{cases}{
\frac{\left(\ph-1\right)!!}{\left(\ph+2q\right)!!}J_{++}\aaa^{2q+1}& $q\geq0$\\
J_{++}\aaa^{\dag2\left|q\right|-1}\frac{(\ph)!!}{\left(\ph+2|q|-1\right)!!}  & $q < 0$
}\end{cases}
.\end{equation*}
In the above, $n!!=n\times(n-2)!!$ is the double factorial. To show that these operators are indeed conserved, first note that an operator $J$ evolves under Eq.~(\ref{eq:Two-Photon}) in the Heisenberg picture, i.e.,
\vbe\label{eq:hei}
\td{J}=\half\kappa_{2\pht}\left(\left[\a^{\star2}\aaa^{2} - \a^2\aaa^{\dag2} ,J\right]+2\aaa^{\dag2} J\aaa^{2}-\aaa^{\dag2}\aaa^{2}J-J\aaa^{\dag2}\aaa^{2}\right)
.\vee
For the case of Eq.~(\ref{eq:j00}), it is easy to see that $\td{J}_{++}=0$ since the two-photon system preserves photon number parity and $J_{++}$ is merely the positive parity projector. The off-diagonal quantity from Eq.~(\ref{eq:j01}) is an extension of $J_{+-}^{(0)}$, the corresponding conserved quantity for the \textit{non-driven} ($\a=0$) dissipative two-photon process (first calculated in~\cite{Simaan1978}; see also~\cite{Albert-Jiang-2013}). Each $J_{+-}^{(q)}$ term in the sum for $J_{+-}$ evolves under Eq.~(\ref{eq:hei}) as

\begin{equation*}
\td{J}_{+-}^{(q)}=\half\k_{2\pht}\left(2q+1\right)\left[\a^{2}J_{+-}^{\left(q-1\right)} -\a^{\star2}J_{+-}^{\left(q+1\right)}-2qJ_{+-}^{\left(q\right)}\right]
.\end{equation*}
The above equations of motion for $J_{+-}^{(q)}$ mimic the recurrence relation
\begin{equation*}
\modas\left[I_{q-1}\left(\modas\right)-I_{q+1}\left(\modas\right)\right]+2qI_{q}\left(\modas\right)=0
\end{equation*}
satisfied by the Bessel functions in $J_{+-}$ and both can be used to verify that $J_{+-}$ is indeed conserved. The square root in front of the sum for $J_{+-}$ is chosen such that $\tr{J_{+-}^\dag|\CCC_\a^+\ke\langle\CCC_\a^-|}=1$, which can be verified using
\vbe\label{eq:jq}
\cmb J_{+-}^{(q)\dag}\cp=\sqrt{\frac{2\modas}{\sinh\left(2\modas\right)}}I_{q}\left(\modas\right)e^{i\t_\a(2q+1)}
\vee
as well as the identity (see Eq. (5.8.6.2) from~\cite{prudnikov})
\vbe\label{eq:iden}
\sum_{q=-\infty}^{\infty}\frac{\left(-1\right)^{q}}{2q+1}I_{q}(\modas)I_{q}(\modas)=\frac{\sinh\left(2\modas\right)}{2\modas}
.\vee

\subsection{Asymptotic state for an initial coherent state of the two-photon process}

The conserved quantities $\{J_{++},J_{+-}\}$ are sufficient to calculate the population $c_{++}=\cpb\rinf\cp$ and coherence $c_{+-}=\cmb\rinf\cp$ of the asymptotic state for any initial state $\rin$. Letting $\rin=|\b\ke\langle\b|$ with $\b=\modb e^{i \t_\b}$, the respective terms are
\vba
c_{++} &= \tr{J_{++}^\dag\rin}=\half (1+e^{-2\modbs}) \label{eq:cpp}\\\label{eq:cpm}
c_{+-} &= \tr{J_{+-}^\dag\rin}=\frac{i\a\b^{\star}e^{-\modbs}}{\sqrt{2\sinh\left(2\modas\right)}}\int_{\phi=0}^{\pi}d\phi e^{-i\phi}I_{0}\left(\left|\a^{2}-\b^2 e^{2i\phi}\right|\right)
.\vea
Eq.~(\ref{eq:cpp}) is the same simple result as the non-driven case (e.g. Eq. (3.22) in~\cite{Albert-Jiang-2013}). To derive Eq.~(\ref{eq:cpm}), we first apply Eq.~(\ref{eq:j01}) to obtain the sum
\begin{equation}\label{eq:int}
c_{+-}=\frac{\sqrt{2}\a\b^{\star}e^{-\modbs}}{\sqrt{\sinh\left(2\modas\right)}}\sum_{q=-\infty}^{\infty}\frac{\left(-1\right)^{q}}{2q+1}I_{q}\left(\modas\right)I_{q}\left(\modbs\right) e^{i2q(\t_{\a}-\t_{\b})}
.\end{equation}
This sum is convergent because the sum without the $2q+1$ term is an addition theorem for $I_q$ (Eq. (5.8.7.2) from~\cite{prudnikov}). To put the above into integral form, we use the identity
(derivable from the addition theorem)
\begin{equation*}
I_{q}\left(\modas\right)I_{q}\left(\modbs\right)
=\frac{1}{2\pi}\int_{\phi=0}^{2\pi}d\phi e^{iq\left(\phi+\pi\right)}I_{0}\left(\left|\modas-\modbs e^{i\phi}\right|\right)
.\end{equation*}
Plugging in the above identity into Eq.~(\ref{eq:int}), interchanging the sum and integral (possible because of convergence), evaluating the sum (which is a simple Fourier series), and performing a change of variables obtains Eq.~(\ref{eq:cpm}).

When $\a=0$, Eq.~(\ref{eq:cpm}) reduces to Eq. (14) from~\cite{Simaan1978}. Assuming real $\a$ and using Eq. (5.8.1.15) from~\cite{prudnikov}, one can calculate limits for large $\modbs$ along the real and imaginary axes in phase space:
$$
\lim_{\b\rightarrow\infty} c_{+-} = \frac{1}{2}\frac{\textrm{erf}(\sqrt{2}\moda)}{\sqrt{1-e^{-4\modas}}}
\stackrel{_{\moda\rightarrow\infty}}{\longrightarrow}\half
\qquad\textrm{ and }\qquad
\lim_{\b\rightarrow i\infty} c_{+-} = -i\frac{1}{2}\frac{\textrm{erfi}(\sqrt{2}\moda)}{\sqrt{e^{4\modas}-1}}
\stackrel{_{\moda\rightarrow\infty}}{\longrightarrow}0
,$$
where erf$(.)$ and erfi$(.)$ are the error function and imaginary error function, respectively. Both limits analytically corroborate Fig.~\ref{fig:charac_2ph} and show that the two-photon system is similar to a classical double-well system in the combined large $\a,\b$ regime.

\subsection{Influence of dephasing on the two-photon process}

Equation~(\ref{eq:rinf}) implies that while the states $\ket{\CCC_\a^\pm}$ define the basis of our logical qubit, the expectation values of the conserved quantities determine the state of the qubit (or equivalently its Bloch vector). Now let's consider adding the photon dephasing dynamics $\k_\phi\DDD\left[\ph\right]$ to Eq.~(\ref{eq:Two-Photon}) and estimate what would happen to the qubit basis elements and more importantly the conserved quantities (determining the effect on the encoded information).

\begin{figure}[h]
\begin{center}
\includegraphics[width=\textwidth]{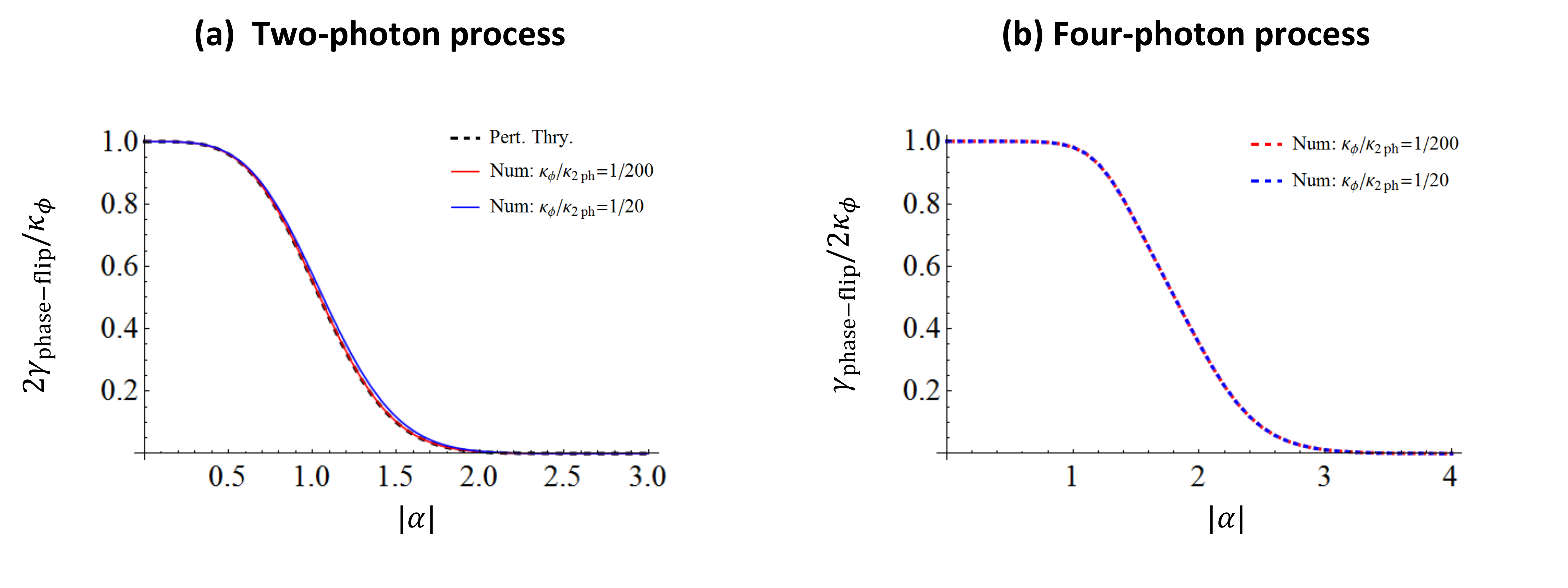}
\end{center}
\caption{{\bf (a)} Plot versus $|\a|$ of the eigenvalue $\g_{\textrm{\scriptsize phase-flip}}$ (scaled by $\k_\phi/2$) of the evolution operator of Eq.~(\ref{eq:deph}) associated with the decay of $J_{+-}$ ($J_{+-}(t)=J_{+-}(0)e^{-\gamma_{\textrm{\tiny{phase-flip}}}t}$). The plot includes the analytical estimate from Eq.~(\ref{eq:pert}) as well as two numerical plots for various $\k_\phi/\k_{2\pht}$. One can see that the eigenvalue exponentially converges to zero with increasing the photon number in the cat state $|\a|^2$. {\bf (b)} Similar plot for Eq.~(\ref{eq:Four-Photon}) with the addition of $\k_\phi\DDD\left[\ph\right]$, i.e., the eigenvalue of the evolution operator associated to the decay of $J_{02}$ encoding the coherence term $|\CCC_\a^{(0\modf)}\ke\langle\CCC_\a^{(2\modf)}|$ of the four-photon process qubit. The phase-flip rate is now scaled by $2\k_\phi$ which represents the rate for the case of $\a=0$ .} 
\label{fig:app}
\end{figure}

Since dephasing preserves parity, the positive parity projector $J_{++}$ remains conserved and the corresponding population of the cat-state  $c_{++}$ thus remains unchanged. The quantity representing the coherence ($J_{+-}$) to first order decays exponentially at a rate proportional to  $\k_\phi$. Noting that the population of the states $\ket{\pm_Z}=\ket{\CCC_\alpha^\pm}$ are conserved, this means that photon dephasing induces only phase-flip errors on our logical qubit. However, this phase-flip rate is itself exponentially suppressed with increasing the number of photons in the cat state $|\a|^2$. To see this, we evaluate the first-order perturbative correction due to dephasing on the asymptotic manifold. Since $\cp\cmb$ and $J_{+-}$ are right and left eigenvectors of the super-operator from Eq.~(\ref{eq:Two-Photon}) and since dephasing preserves parity, the first order decay rate $\g_{\textrm{\scriptsize phase-flip}}$ is
\begin{equation*}
\g_{\textrm{\scriptsize phase-flip}}=\k_{\phi}\tr{J_{+-}^{\dag}\mathcal{D}\left[\ph\right]\cp\cmb}=\k_{\phi}\cmb\mathcal{D}\left[\ph\right]J_{+-}^{\dag}\cp
.\end{equation*}
In the above, we have re-arranged for the adjoint of $\DDD$ to act on $J_{+-}$ instead of $\cp\cmb$ and used $\DDD^\dag\left[\ph\right]=\DDD\left[\ph\right]$ because $\ph$ is Hermitian. Since $J_{+-}^{(q)}$ consist of matrix elements $|2n\ke\langle 2n+1+2q|$ for $n=0,1,...$, each term in the sum for $J_{+-}$ has the simple equation of motion
\begin{equation*}
\mathcal{D}\left[\ph\right]J_{+-}^{\dag(q)}=-\half\k_{\phi}(2q+1)^{2}J_{+-}^{\dag(q)}
.\end{equation*}
The subsequent evaluation of the trace and sum results in the rate 
\begin{equation}\label{eq:pert}
\g_{\textrm{\scriptsize phase-flip}}=-\k_{\phi}\frac{\modas}{\sinh(2\modas)}
\end{equation}
given in Sec.~\ref{ssec:twoPh}. We have numerically confirmed [Fig.~\ref{fig:app}(a)] that this is indeed the first-order correction to the asymptotic manifold. In the Figure, we plot versus $\moda$ the magnitude of the eigenvalue of the evolution operator from Eq.~(\ref{eq:deph}) associated with the decay rate of $J_{+-}$ (which is precisely the phase-flip rate $\gamma_{\textrm{\scriptsize phase-flip}}$). For small values of $\k_\phi/\k_{2\textrm{\scriptsize ph}}$, the numerical result approaches our analytical estimate.

It is worth noting that under the effect of dephasing, the cat-states that comprise the logical qubit basis elements will acquire a small random phase ($\ket{\CCC_\alpha^\pm}$ becomes $\ket{\CCC_{\alpha e^{i\phi}}^\pm}$ where $\phi$ is a small random phase). Indeed, as an ensemble-averaged result, one can observe that each of the two-dimensional Gaussian peaks that represent the cat state in the phase space slightly smear. However, this smearing merely changes the structure of our qubit basis elements and does not affect the encoded quantum information (represented by $J_{++}$ and $J_{+-}$). 
\subsection{Asymptotic behavior of the four-photon process}

The asymptotic manifold of the four-photon process from Eq.~(\ref{eq:Four-Photon}) is given by density matrices defined on the four-dimensional Hilbert space spanned by $\{|\CCC_\a^{(\m\modf)}\ke\}$ (with $\m=0,1,2,3$). By tracking the parity, we restrict the dynamics to the Hilbert space spanned by $\{\ket{\CCC_\a^{(0\modf)}},\ket{\CCC_\a^{(2\modf)}}\}$ comprising our logical qubit's basis. The corresponding conserved quantity for the populations of $\ket{\CCC_\a^{(0\modf)}}$ and $\ket{\CCC_\a^{(2\modf)}}$ is once again identical to the non-driven case~\cite{Albert-Jiang-2013}, $J_{00}=\sum_{n=0}^{\infty}\ket{4n}\bra{4n}$. While an analytical expression for the other conserved quantity $J_{02}$ remains to be found, here we provide a numerical analysis of the influence of the photon dephasing on the four-photon process. 

Fig.~\ref{fig:app}(b) shows a plot similar to Fig.~\ref{fig:app}(a), but now for $\g_{\textrm{\scriptsize phase-flip}}$ of the logical qubit of the four-photon process. With the exception of a slight delay in the exponential suppression of the induced phase-flip rate, one observes that this suppression is almost identical to the case of the two-photon process.




\section*{References}
\begin{thebibliography}{10}

\bibitem{Leghtas-al-PRL-2013}
Z.~Leghtas, G.~Kirchmair, B.~Vlastakis, R.J. Schoelkopf, M.H. Devoret, and
  M.~Mirrahimi.
\newblock Hardware-efficient autonomous quantum memory protection.
\newblock {\em Phys. Rev. Lett.}, 111, 2013.

\bibitem{Shor-QEC}
P.~Shor.
\newblock Scheme for reducing decoherence in quantum memory.
\newblock {\em Phys. Rev. A}, 52:2493--2496, 1995.

\bibitem{Steane-PRL_1996}
A.~Steane.
\newblock Error correcting codes in quantum theory.
\newblock {\em Phys. Rev. Lett}, 77(5), 1996.

\bibitem{nielsen-chang-book}
M.A. Nielsen and I.L. Chuang.
\newblock {\em Quantum Computation and Quantum Information}.
\newblock Cambridge University Press, 2000.

\bibitem{Leghtas-al-PRA-2013}
Z.~Leghtas, G.~Kirchmair, B.~Vlastakis, M.H. Devoret, R.J. Schoelkopf, and
  M.~Mirrahimi.
\newblock Deterministic protocol for mapping a qubit to coherent state
  superpositions in a cavity.
\newblock {\em Phys. Rev. A}, 87, 2013.

\bibitem{schuster-nature07}
D.I. Schuster, A.A. Houck, J.A Schreier, A.~Wallraff, J.M. Gambetta, A.~Blais,
  L.~Frunzio, J.~Majer, B.~Johnson, M.H. Devoret, S.M. Girvin, and R.~J.
  Schoelkopf.
\newblock Resolving photon number states in a superconducting circuit.
\newblock {\em Nature}, 445:515--518, 2007.

\bibitem{Carmichael-Wolinsky-1988}
H.J. Carmichael and M.~Wolinsky.
\newblock Quantum noise in the parametric oscillator: From squeezed states to
  coherent-state superpositions.
\newblock {\em Phys. Rev. Lett.}, 60:1836--1839, 1988.

\bibitem{Krippner-et-al-1994}
L.~Krippner, W.J. Munro, and M.D. Reid.
\newblock Transient macroscopic quantum superposition states in degenerate
  parametric oscillation: Calculations in the large-quantum-noise limit using
  the positive {P} representation.
\newblock {\em Phys. Rev. A}, 50:4330--4338, 1994.

\bibitem{Hach-Gerry-1994}
E.~Hach III and C.C. Gerry.
\newblock Generation of mixtures of {S}chr\"{o}dinger-cat states from a
  competitive two-photon process.
\newblock {\em Phys. Rev. A}, 49:490--498, 1994.

\bibitem{Gilles-et-al-1994}
L.~Gilles, B.M. Garraway, and P.L. Knight.
\newblock Generation of nonclassical light by dissipative two-photon processes.
\newblock {\em Phys. Rev. A}, 49:2785--2799, 1994.

\bibitem{Everitt-et-al-2013}
M.J. Everitt, T.P. Spiller, G.J. Milburn, R.D. Wilson, and A.M. Zagoskin.
\newblock Cool for {C}ats.
\newblock 2012.
\newblock arXiv:1212.4795.

\bibitem{Gilchrist-et-al-2004}
A.~Gilchrist, K.~Nemoto, W.J. Munro, T.C. Ralph, S.~Glancy, S.L. Braunstein,
  and G.J. Milburn.
\newblock Schr\"{o}dinger cats and their power for quantum information
  processing.
\newblock {\em J. Opt. B: Quantum Semiclass. Opt.}, 6(8), 2004.

\bibitem{Facchi-Pascazio-2002}
P.~Facchi and S.~Pascazio.
\newblock Quantum {Z}eno subspaces.
\newblock {\em Phys. Rev. Lett.}, 89(8):080401, 2002.

\bibitem{raimond-et-al-2010}
J.-M. Raimond, C.~Sayrin, S.~Gleyzes, I.~Dotsenko, M.~Brune, S.~Haroche,
  P.~Facchi, and S.~Pascazio.
\newblock Phase space tweezers for tailoring cavity fields by quantum zeno
  dynamics.
\newblock {\em Phys. Rev. Lett.}, 105:213601, 2010.

\bibitem{Raimond-et-al-2012}
J.M. Raimond, P.~Facchi, B.~Peaudecerf, S.~Pascazio, C.~Sayrin, I.~Dotsenko,
  S.~Gleyzes, M.~Brune, and S.~Haroche.
\newblock Quantum {Z}eno dynamics of a field in a cavity.
\newblock {\em Phys. Rev. A}, 86:032120, 2012.

\bibitem{Lidar-et-al-98}
D.A. Lidar, I.L. Chuang, and K.B. Whaley.
\newblock Decoherence-free subspaces for quantum computation.
\newblock {\em Phys. Rev. Lett.}, 81(12):2594--2597, 1998.

\bibitem{Yurke-Stoler-1986}
B.~Yurke and D.~Stoler.
\newblock Generating quantum mechanical superpositions of macroscopically
  distinguishable states via amplitude dispersion.
\newblock {\em Phys. Rev. Lett.}, 57:13--16, 1986.

\bibitem{Kirchmair-al-Nature_2013}
G.~Kirchmair, B.~Vlastakis, Z.~Leghtas, S.E. Nigg, H.~Paik, E.~Ginossar,
  M.~Mirrahimi, L.~Frunzio, S.M. Girvin, and R.J. Schoelkopf.
\newblock Observation of quantum state collapse and revival due to the
  single-photon {K}err effect.
\newblock {\em Nature}, 495:205, 2013.

\bibitem{Kumar-Divincenzo-2010}
S.~Kumar and D.P. Di{V}incenzo.
\newblock Exploiting {K}err cross nonlinearity in circuit quantum
  electrodynamics for nondemolition measurements.
\newblock {\em Phys. Rev. B}, 82:014512, 2010.

\bibitem{Vlastakis-et-al-Science_2013}
B.~Vlastakis, G.~Kirchmair, Z.~Leghtas, S.E. Nigg, L.~Frunzio, S.M. Girvin,
  M.~Mirrahimi, M.H. Devoret, and R.J. Schoelkopf.
\newblock Deterministically encoding quantum information using 100-photon
  {S}chr\"{o}dinger cat states.
\newblock {\em Science}, 342:607--610, 2013.

\bibitem{haroche-raimond:book06}
S.~Haroche and J.M. Raimond.
\newblock {\em Exploring the Quantum: Atoms, Cavities and Photons.}
\newblock Oxford University Press, 2006.

\bibitem{haroche-et-al-2007}
S.~Haroche, M.~Brune, and J.M Raimond.
\newblock Measuring the photon number parity in a cavity: from light quantum
  jumps to the tomography of non-classical field states.
\newblock {\em Journal of Modern Optics}, 54:2101, 2007.

\bibitem{Sun-et-al-2013}
L.~Sun, A.~Petrenko, Z.~Leghtas, B.~Vlastakis, G.~Kirchmair, K.~Sliwa,
  A.~Narla, M.~Hatridge, S.~Shankar, J.~Blumoff, L.~Frunzio, M.~Mirrahimi, M.H.
  Devoret, and R.J. Schoelkopf.
\newblock Tracking photon jumps with repeated quantum non-demolition parity
  measurements.
\newblock 2013.
\newblock submitted, arXiv:1311.2534.

\bibitem{Herviou-Mirrahimi}
L.~Herviou and M.~Mirrahimi.
\newblock Fault-tolerant photon number parity measurement for a quantum
  harmonic oscillator.
\newblock 2013.
\newblock in preparation.

\bibitem{Vijay-et-al-09}
R.~Vijay, M.H. Devoret, and I.~Siddiqi.
\newblock Invited review article: The {J}osephson bifurcation amplifier.
\newblock {\em Rev. Sci. Instrum.}, 80:111101, 2009.

\bibitem{nigg-et-al-2012}
S.E. Nigg, H.~Paik, B.~Vlastakis, G.~Kirchmair, S.~Shankar, L.~Frunzio, M.H.
  Devoret, R.J. Schoelkopf, and S.M. Girvin.
\newblock Black-box superconducting circuit quantization.
\newblock {\em Phys. Rev. Lett.}, 108:240502, Jun 2012.

\bibitem{Drummond-et-al-1981}
P.D. Drummond, K.J. McNeil, and D.F. Walls.
\newblock Non-equilibrium transitions in sub/second harmonic generation {II.}
  {Q}uantum theory.
\newblock {\em Opt. Acta}, 28:211--225, 1981.

\bibitem{Bergeal-et-al-2010}
N.~Bergeal, R.~Vijay, V.~E. Manucharyan, I.~Siddiqi, R.J. Schoelkopf, S.M.
  Girvin, and M.H. Devoret.
\newblock Analog information processing at the quantum limit with a {J}osephson
  ring modulator.
\newblock {\em Nature Physics}, 6:296--302, 2010.

\bibitem{Roch-et-al-12}
N.~Roch, E.~Flurin, F.~Nguyen, P.~Morfin, P.~Campagne-Ibarcq, M.~H. Devoret,
  and B.~Huard.
\newblock Widely tunable, non-degenerate three-wave mixing microwave device
  operating near the quantum limit.
\newblock {\em Phys. Rev. Lett.}, 108:147701, 2012.

\bibitem{Schackert-thesis}
F.~Schackert.
\newblock {\em A practical quantum-limited parametric amplifier based on the
  {J}osephson ring modulator}.
\newblock PhD thesis, Yale University, 2013.

\bibitem{Albert-Jiang-2013}
V.V. Albert and L.~Jiang.
\newblock Symmetries and conserved quantities in {L}indblad master equations.
\newblock 2013.
\newblock arXiv:1310.1523.

\bibitem{Simaan1978}
H.D. Simaan and R.~Loudon.
\newblock Off-diagonal density matrix for single-beam two-photon absorbed
  light.
\newblock {\em J. Phys. A: Math. Gen.}, 8:539, 1975.

\bibitem{prudnikov}
A.P. Prudnikov, Y.A. Brychkov, and O.I. Marichev.
\newblock {\em Integrals and Series Vol. 2: Special Functions}.
\newblock {Gordon and Breach Science Publishers}, 1st edition, 1992.

\end{thebibliography}
\end{document}